\documentclass[structabstract]{aa}  
\usepackage[tight]{subfigure}
\usepackage{graphicx}
\usepackage{txfonts}
\usepackage{natbib}
\bibpunct{(}{)}{;}{a}{}{,}

\usepackage{color}

\begin{document}
   \title{Three-dimensional simulations of the interaction between Type Ia supernova ejecta and their main sequence companions}


   \author{Z.~W.~Liu
          \inst{1,2,3,4},
           R.~Pakmor
          \inst{5,4},
           F.~K.~R{\"o}pke
          \inst{6,4},
           P.~Edelmann
          \inst{4},
           B.~Wang
          \inst{1,2},
           M.~Kromer
          \inst{4},
           W.~Hillebrandt
          \inst{4}
          \and
           Z.~W.~Han
          \inst{1,2}  }

   \institute{National Astronomical Observatories/Yunnan Observatory, Chinese Academy of Sciences, Kunming 650011, P.R. China\
         \and
              Key Laboratory for the Structure and Evolution of Celestial Objects, Chinese Academy of Sciences, Kunming 650011, P.R. China\
         \and
              Graduate University of Chinese Academy of Sciences, Beijing 100049, P.R. China\
         \and
              Max-Planck-Institute f\"ur Astrophysik, Karl-Schwarzschild-Str. 1, 85741 Garching, Germany\
         \and
              Heidelberger Institut f\"ur Theoretische Studien, Schloss-Wolfsbrunnenweg 35, 69118 Heidelberg, Germany\
         \and
              Institut f\"ur Theoretische Physik und Astrophysik,
              Universit\"at W\"urzburg, Emil-Fischer-Str.~31, 97074 W\"urzburg, Germany\\
             \email{zwliu@mpa-garching.mpg.de}  }

   \date{Received April 05, 2012; accepted September 18, 2012}

   \abstract
   {The identity of the progenitor systems of Type Ia supernovae
     (SNe~Ia) is still uncertain.  In the single-degenerate scenario,
     the interaction between the supernova blast wave and the outer
     layers of a main sequence companion star strips off hydrogen-rich
     material which is then mixed into the ejecta. Strong
     contamination of the supernova ejecta with stripped material
     could lead to a conflict with observations of SNe~Ia. This
     constrains the single-degenerate progenitor model.}
   {In this work, our previous simulations based on simplified progenitor donor stars have been updated by adopting more
     realistic progenitor-system models that result from fully
     detailed, state-of-the-art binary evolution calculations.}
   { We use Eggleton's stellar evolution code including the optically
     thick accretion wind model and taking into
     account the possibility of the effects of accretion disk
     instabilities to obtain realistic models of companion stars for
     different progenitor systems. The impact of the supernova blast
     wave on these companion stars is followed in three-dimensional
     hydrodynamic simulations employing the smoothed particle
     hydrodynamics (SPH) code GADGET3.}
   {For a suite of main sequence companions, we find that the
     mass of the material stripped from the companions range from
     0.11~$M_\odot$ to 0.18~$M_\odot$. The kick velocity delivered
     by the impact is between 51~km~s$^{-1}$ and 105~km~s$^{-1}$. We
     find that the stripped mass and kick velocity depend on the
     ratio of the orbital separation to the radius of a companion,
     $a_{\mathrm{f}}/R$. They can be fitted in good
     approximation by a power law  for a given companion model. However, we do not
     find a single power law relation holding for different companion
     models. This implies that the structure of the companion star is
     also important for the amount of stripped material.}
   {With more realistic companion star models than those employed in
     previous studies, our simulations show that the hydrogen masses
     stripped from companions are inconsistent with the best
     observational limits ($\leqslant 0.01 M_{\odot}$) derived from SN
     Ia nebular spectra. However, a rigorous forward modeling from the
     results of impact simulations with radiation transfer is required
     to reliably predict observable signatures of the stripped
     hydrogen and to conclusively assess the viability of the
     considered SN~Ia progenitor scenario.}

   \keywords{stars: supernovae: general --
                hydrodynamics--binaries: close
               }

   \authorrunning{Z. W. Liu et al.}

   \titlerunning{3D simulations of interaction between SN~Ia ejecta and their MS companions}

   \maketitle
%

\section{Introduction}

Type Ia supernovae (SNe~Ia) play a fundamental role in astrophysics.
They are one of the most powerful tools in cosmology due to their extreme luminosities
and high homogeneity. Based on an empirical relation between the light
curve shape and the peak luminosity \citep{Phil93, Phil99}, they can
be used as accurate cosmic distance indicators. This led to the
discovery of the accelerating expansion of the Universe
\citep{Ries98, Perl99, Leib08}.  However, despite recent progress on
both, the theoretical and observational side, the nature of SN~Ia
progenitors and the physics of the explosion mechanisms
are still unclear \citep{Hill00}.

It is widely accepted that SNe~Ia arise from a mass-accreting white
dwarf (CO WD) undergoing a thermonuclear explosion
(for a review see \citealt{Hoyl60, Finz67, Nomo82}).  At present, the most general classification
of progenitor scenarios are \emph{single-degenerate\/} and
\emph{double-degenerate\/} models. In the single degenerate (SD)
scenario, a white dwarf (WD) accretes hydrogen-rich material from its
non-degenerate companion, where the companion star could be a
main-sequence (MS) star (WD + MS channel), a slightly evolved subgiant
star or a red giant (RG) star. When the mass of the WD approaches the
Chandrasekhar mass limit, it explodes as a SN~Ia \citep[e.g.][]{Whel73, Hach96, Han04,Wang10b}.  
In the double degenerate (DD) scenario, two CO WDs spiral
in and merge due to gravitational wave radiation, resulting in a
single object with a mass above the Chandrasekhar limit, which then
may explode as SN~Ia \citep{Iben84, Webb84}.

However, neither the SD nor the DD models can yet be clearly favored
from observation or theory \citep{Maoz11}. In the SD case, only a fairly
narrow range in the accretion rate above $10^{\mathrm{-7}}\ M_{\mathrm{\odot}}\,\mathrm{yr^{-1}}$ 
is allowed in order to attain stable hydrogen
burning on the surface of the WD, avoiding a nova explosion. This makes it
difficult to explain the observed nearby SN~Ia rate \citep{Mann05,
  Maoz11}. Recent observations have identified some DD binaries
\citep{Nele05, Geie07}. However, only few DD systems have been found whose
orbital period is short enough to merge in a Hubble time \citep{Geie10, Rodr10}. In none of
them, the combined mass exceeds
the Chandrasekhar limit.  In addition, the DD channel has been
suggested to lead to an accretion-induced collapse rather than an SN
Ia \citep{Nomo85, Saio98}. Recently, however, \citet{Pakm10, Pakm11b}
found that violent mergers of equal-mass white dwarfs with masses
$\sim 0.9\,M_\odot$ can directly trigger thermonuclear explosions which
resemble sub-luminous 1991bg-like SNe~Ia.  The violent merger of two CO WDs with masses of 
$0.9\,M_{\mathrm{\odot}}$ and $1.1\,M_{\mathrm{\odot}}$ can lead to events that
reproduce observational characteristics of normal SNe~Ia
\citep{Pakm12b}.

With detailed binary population synthesis (BPS) calculations, some
research groups investigated the delay-time distributions (DTD) and
birthrates of SNe~Ia for different formation channels. Theoretical
predictions are compared with
observations to constrain the progenitor
systems of SNe~Ia and their forming scenarios \citep{Ruit09, Mann09, Wang09, Wang10a,Maoz10a, Maoz10b}.  
However, there is no evidence that the SD
scenario alone explains the 
shape of the observed delay-time distribution, while this may be
possible for the DD scenario or a
combination of both channels \citep{Menn10}.

The obvious difference between the SD and DD scenario is that the
companion stars in the SD channel will survive and may affect the
observational display of the explosion \citep{Mari00, Pakm08}.
In contrast, no stellar remnant exists in the case of
the merger of two WDs in the DD channel. It is a promising approach to
constrain the nature of SN~Ia progenitors by directly searching for
the surviving companion stars in remnants of SNe~Ia.  A prominent
example is Tycho Brahe's supernova (SN 1572) for which \citet{Ruiz04} presented a survey of the central
region of its remnant, around the position of the explosion. They
identified a subgiant star (Tycho G), similar to the Sun in surface
temperature but with very low surface gravity, which moves at more
than three times the mean velocity of the stars in the neighborhood
and suggested that this might be the surviving companion of SN 1572.
However, the study of \citet{Kerz09} casts some doubts on this
identification.  Recently, \citet{Scha12} analyzed the center of the LMC
SN~Ia remnant SNR 0509-67.5 which does not contain any stars down to the
observational limit. Thus, all possible companion star types except
white dwarfs can be excluded for this object.

In the SD scenario, hydrogen-rich or helium-rich circum-stellar
material (CSM) is expected to exist around SNe~Ia as the result of
mass transfer from the companion, as well as WD winds \citep{Nomo82,Hach99b}. 
Therefore, another indirect way of identifying SN~Ia
progenitor systems is to search for the material transferred to the
accreting white dwarf in the CSM \citep{Pata07}. Following this
approach \citet{Pata07} reported detection of such CSM. They
suggested that the SD system consisting of a WD and RG is the favored
progenitor for SN 2006X \citep{Pata07}. Moreover, \citet{Ster11} 
showed that the velocity structure of absorbing material along the line of sight to 35 SNe Ia
tends to be blueshifted. Thus, they concluded that many SNe Ia in nearby spiral 
galaxies may originate in SD systems \citep{Ster11}. Note, however, that
abundant CSM is in
conflict with the missing radio signal for other events, e.g.\ SN 2011fe \citep{Chom12,Hore12}.

SNe~Ia are characterized by the lack of hydrogen in their
spectra. This is explained naturally in the DD progenitor
scenario. However, in the SD scenario, the companion star is typically
hydrogen-rich. Thus, the impact of a SN~Ia explosion will strip off
hydrogen-rich material from the companion and mix it into the
ejecta, which may be in conflict with observations if the amount of
hydrogen is large enough to be observable. Thus, we may be able to
identify the progenitors of SNe~Ia based on the exact amount of
hydrogen stripped from the surface of the companion star. So
far, there is no direct detection of the stripped
hydrogen. Based on high-resolution spectroscopy of the two
extragalactic SNe~Ia SN 2005am and SN 2005cf, \citet{Leon07} estimated
an upper limit of $0.01\,M_{\odot}$ applying the model described by
\citet{Matt05}.

\citet[][]{Mari00} (hereafter~M00) explored  the impact of SN~Ia ejecta on
a variety of binary companions (MS, subgiants, RGs) in the SD
formation channel with two dimensional Eulerian
hydrodynamics simulations. They found that the supernova explosion can strip
$0.15\,M_{\odot}$ to $0.17\,M_{\odot}$ of material from the surface of MS and
subgiant companions, while RGs lose almost their entire envelope
in the impact. However, they
ignored the effect of the mass transfer phase on the structure of the
companion star when they constructed their initial model. To investigate how the mass transfer changes the mass stripping by
supernova explosions, \citet{Meng07} used an analytical method to
approximate the mass loss in the impact. They found a lower limit
of 0.035\,$M_{\odot}$ for the stripped mass, but their analytic method
was based on an oversimplified description of the interaction physics.

An updated study has been presented
by \citet[hereafter~PRWH08]{Pakm08} based on three-dimensional (3D)
smoothed particle hydrodynamics (SPH) simulations. By mimicking
the models of \citet{Ivan04} in their setups, they found stripped
masses in the range
from $0.01\,M_{\mathrm{\odot}}$ to $0.06\,M_{\mathrm{\odot}}$, which is very
close to the  observational limits of $\sim 0.01\,M_{\odot}$.
Therefore, they concluded that the SD scenario remains a
valid possibility for SN~Ia progenitors. Although mass-loss from the
companion star was included in their study, it was modeled by removal
of material from a main sequence star with a constant rate in
single-star evolution code. This is an oversimplification and we therefore
aim at reexamining the impact of supernova ejecta on a companion star
that was modeled consistently in a detailed binary evolution
calculation.

Very recently, Pan et al. (2012) studied the impact of SN~Ia ejecta on
binary companions in the SD scenario with the Eulerian hydrodynamics
code FLASH for MS, RG
and He star companions. They were able to quantify the amount of
contamination with explosion ashes on the companion star by the supernova ejecta in their
simulations which might help to identify a companion star even a long
time after the explosion.

\begin{table*}
\caption{Main-sequence companion star models.} \label{table:1}
\centering
\begin{tabular}{c c c c l l l l l l}     
\hline\hline
 Model \tablefootmark{1}   & $M_{\mathrm{wd}}$  & $M_{\mathrm{c,i}}$ & $M_{\mathrm{c,f}}$   &\ \ $P_{\mathrm{f}}$ &\ \ \ $a_{\mathrm{f}}$   &   $a_{\mathrm{f}}/R_2$  &\ \ $M_{\mathrm{stripped}}$ & \ \ $v_{\mathrm{kick}}$    &SNe~Ia \tablefootmark{2}   \\
 &[$M_{\mathrm{\odot}}$]&[$M_{\mathrm{\odot}}$]&[$M_{\mathrm{\odot}}$] &[$d$] & [$10^{11}$cm]&  &\ \ [$M_{\mathrm{\odot}}$]  &[$\mathrm{km\ s^{-1}}$]&  \\
\hline
   ms\_20a & 0.70 & 2.00 & 0.74 &\ \ 0.98 &\ \ \  3.72 & 3.06 &\  \ 0.181 &\ \ 51.01 & disk instability\\
   ms\_20b & 0.90 & 2.00 & 1.17 &\ \ 0.46 &\ \ \  2.40 & 2.75 &\  \ 0.105 &\ \ 58.78 & disk instability\\
   ms\_22a & 0.80 & 2.20 & 1.21 &\ \ 0.29 &\ \ \  1.77 & 2.72 &\  \ 0.173 &\ \ 105.29& weak H-shell flash\\
   ms\_20c & 0.80 & 2.00 & 1.22 &\ \ 0.56 &\ \ \  2.74 & 2.71 &\  \ 0.171 &\ \ 64.32 & weak H-shell flash\\
   ms\_24a & 0.90 & 2.40 & 1.40 &\ \ 0.33 &\ \ \  1.95 & 2.63 &\  \ 0.172 &\ \ 94.95 & stable H burning\\
   ms\_20d & 1.00 & 2.00 & 1.40 &\ \ 0.57 &\ \ \  2.83 & 2.63 &\  \ 0.113 &\ \ 53.32 & weak H-shell flash\\
   ms\_23a & 0.90 & 2.30 & 1.50 &\ \ 0.35 &\ \ \  2.08 & 2.59 &\  \ 0.162 &\ \ 85.66 & stable H burning\\
   ms\_24b & 1.00 & 2.40 & 1.88 &\ \ 0.37 &\ \ \  2.25 & 2.46 &\  \ 0.116 &\ \ 66.91 & stable H burning\\
   ms\_28a & 1.10 & 2.80 & 2.00 &\ \ 0.33 &\ \ \  2.11 & 2.43 &\  \ 0.159 &\ \ 84.50 & optically thick wind\\
   ms\_30a & 1.20 & 3.00 & 2.45 &\ \ 0.44 &\ \ \  2.64 & 2.33 &\  \ 0.141 &\ \ 65.34 & optically thick wind\\
\hline
\end{tabular}
\tablefoot{$M_{\mathrm{wd}}$ and $M_{\mathrm{c,i}}$ are the initial masses of WD and its 
  companion at the beginning of mass transfer, respectively.
  $M_{\mathrm{c,f}}$, $P_{\mathrm{f}}$, $a_{\mathrm{f}}$ and $R_2$ denote the
  final mass of companion star, the orbital period, the binary separation 
  and the radius of secondary at the moment of supernova explosion, 
  respectively. $M_{\mathrm{stripped}}$ and $v_{\mathrm{kick}}$ are the stripped 
  mass and the kick velocity caused by supernova impact.\\
  \tablefoottext{1}{All models have been named with the same way. 
    For example, ms\_20(a,b,c,d), the ``ms'' corresponds to CO WD + MS system,
    the middle number ``20'' means the $M_{\mathrm{c,f}}$
    is $2.0\,M_{\mathrm{\odot}}$. The final alphabet ``a'', ``b'', ``c'' or ``d'' denote 
    the different models with the different $M_{\mathrm{c,f}}$  but the 
    same $M_{\mathrm{c,i}}$.}\\
  \tablefoottext{2}{The WD explodes as an SN~Ia in the disk instability phase, in the optically thick wind phase, in 
    the stable H-shell burning phase and in the weak H-shell flash phase, respectively.}  
}

\end{table*}

Based on the prescription of \citet{Hach99b} for the mass growth of CO
WDs, and including the possibility of the instability of an accretion disk
around the WD on the evolution of binary systems, detailed binary
evolution calculations have been performed for about 2400 close WD
binaries by \citet[hereafter ~WLH10]{Wang10b}.  They confirmed that
WDs in the WD~+~MS channel with an initial mass as low as $0.61\,M_{\odot}$ 
can accrete efficiently and reach the Chandrasekhar mass
limit. Their calculations also showed that the disk instability could
substantially increase the mass-accumulation efficiency for accreting
WDs and cause SNe~Ia to occur in systems with $\leqslant
2\,M_{\odot}$ donor stars. They found that the Galactic SN~Ia birth rate
from the WD~+~MS channel is $\sim 1.8 \times 10^{-3}$ yr$^{-1}$
according to their standard model, which can account for 2/3 of the
observed SNe~Ia.

In this work, we use the same method as WLH10 to carry out
consistent binary evolution calculations for the single degenerate MS
channel of SNe~Ia. With these more realistic companion models, we
expand and update the 3D hydrodynamical simulations performed in
PRWH08 to investigate the interaction of SN~Ia ejecta with MS
companion stars. We then explore how the ejecta structure, the
stripped mass and the kick velocities of the surviving companion
depend on parameters of the progenitor model. Section~\ref{sec:codes} describes the codes and
initial models used in this paper. The SPH impact simulations and
numerical results for ten consistent MS companion models are presented
in Section~\ref{sec:numerical_results}. The comparisons with PRWH08
and some implications of our simulations are discussed in
Section~\ref{sec:discussion}. We summarize and conclude in Section~\ref{sec:conclusion}.


\section{Numerical methods and models}
  \label{sec:codes}

  \subsection{Numerical codes}

   \label{sec:code}

   We use Eggleton's stellar evolution code
   \citep{Eggl71, Eggl72, Eggl73} to follow the detailed binary
   evolution of SD progenitor systems. The latest input physics have
   been implemented by \citet{Wang09, Wang10b}. Roche lobe overflow
   (RLOF) is treated in the code as described by \citet{Han04}. The
   opacity tables in our code have been compiled by \citet{Chen07} from
   \citet{Igle96} and \citet{Alex94}. We use a typical Population I
   composition with hydrogen abundance X = 0.70, helium abundance Y =
   0.28 and metallicity Z = 0.02. We set the ratio of the typical mixing
   length to the local pressure scale height, $\alpha = l/H_{\mathrm{P}}$,
   to 2, and the convective overshooting parameter,
   $\delta_{\mathrm{ov}}$, to 0.12 \citep{Pols97, Schr97}, which roughly
   corresponds to an overshooting length of $\sim 0.25$ pressure scale
   heights ($H_{\mathrm{P}}$).

   \begin{figure*}
   \centering
   \includegraphics[width=12cm]{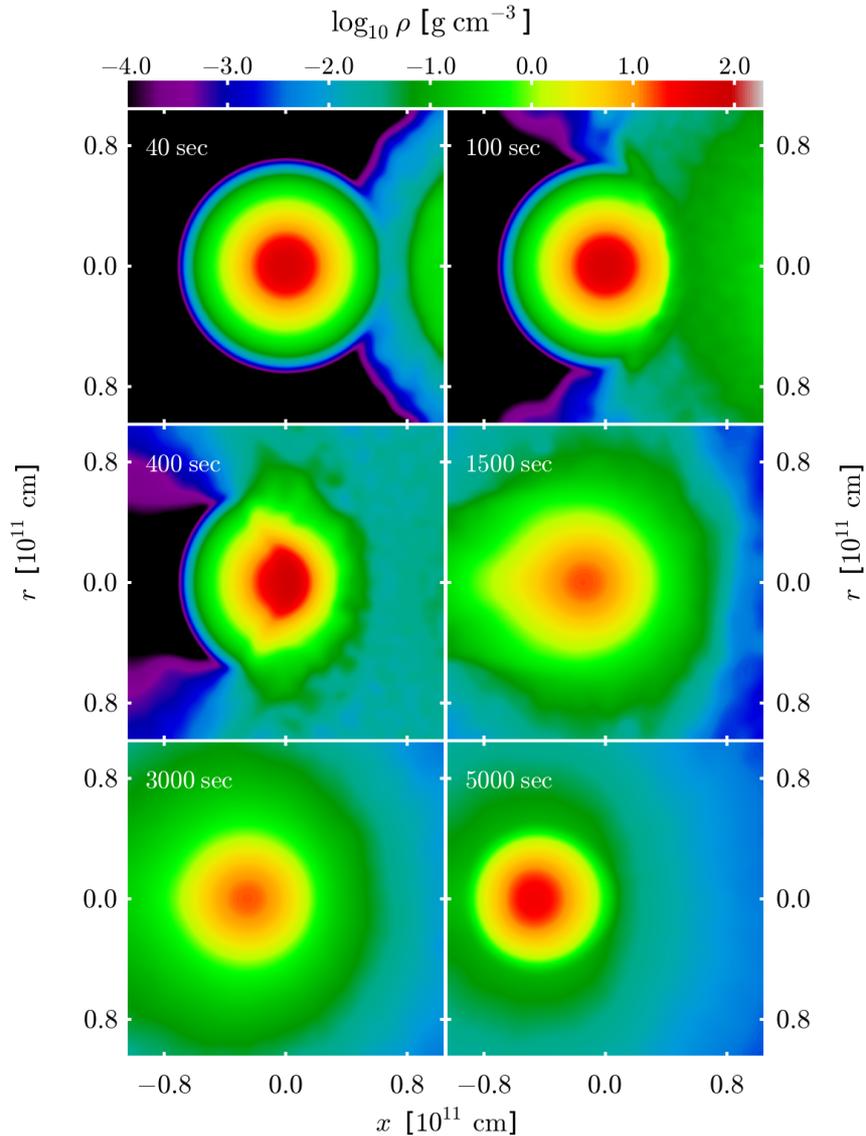}
   \caption{Time evolution of density distribution in the impact simulation with model
     ms\_22a.  For visualization (not for the simulation, though), we use cylindrical coordinates. 
     The radial coordinate is averaged over angle accounting for the intrinsic symmetry of the star. 
     The logarithm of density is color-coded.}
    \label{Fig:dens_ms22a}%
    \end{figure*}

    In this paper, we start our binary evolution calculations from
    WD+MS binary systems.  In a WD+MS binary system, mass transfer
    occurs through RLOF when the companion star fills its Roche
    lobe. If the mass transfer is dynamically stable, the transferred
    material will form an accretion disk surrounding the WD. This
    accretion disk can become thermally unstable when the
    effective temperature in the disk falls below the hydrogen
    ionization temperature $\sim 6500 \mathrm{K}$ \citep{van96, Laso01}.  Here, as a
    possibility, the effect of the instability of the accretion disk
    on the evolution of WD + MS binaries has been included to
    constrain the accretion rate of the WD. If the mass-transfer rate,
    $\left | \dot M_{2} \right |$, is higher than the critical
    mass-transfer rate for a stable accretion disk, $\dot M_{\mathrm{cr,
        disk}}$, we will assume that 
    the WD accretes smoothly at a rate $\dot M_{\mathrm{acc}} = \left |
      \dot M_{2} \right |$.  Otherwise, the disk is unstable and the
    mass-accretion rate of the WD is $\dot M_{\mathrm{acc}} = \left | \dot
      M_{2} \right |/d$, where $d$ is the duty cycle, set to $0.1$ in
    this work.
   
    We do not solve the stellar structure equations for the WD star
    when we construct the structure of the companion star for our
    simulation. Instead, we adopt the prescription of \citet{Hach99b}
    for the mass growth of a CO WD by accretion of hydrogen-rich
    material from a companion.  If the mass-accretion rate of
    the WD, $\dot M_{\mathrm{acc}}$, is above a critical value, $\dot
    M_{\mathrm{cr,WD}}$, we assume that hydrogen burns steadily on the
    surface of the WD and that the hydrogen-rich material is converted
    into helium at a rate $\dot M_{\mathrm{cr,WD}}$, while the unprocessed
    matter is assumed to be lost from the system as an optically thick
    wind at a mass-loss rate $\dot M_{\mathrm{wind}}= \left | \dot M_{2}
    \right | - \dot M_{\mathrm{cr,WD}}$.  The critical mass-accretion rate
    is \citep{Han04}

  \begin{equation}
    \label{equation:1}
     \dot M_{\mathrm{cr, WD}} = 5.3 \times 10^{-7} \frac{(1.7-X)}{X}(M_{\mathrm{WD}}/\mathrm{M_{\odot}}-0.4)\ \mathrm{M_{\mathrm{\odot}}yr^{-1}}.
  \end{equation}
  where $X$ is the hydrogen mass fraction and $M_{\mathrm{WD}}$ is the
  mass of the accreting WD.  \citet{Kato11} pointed out that this WD
  wind case may correspond to the quasi-regular transient supersoft
  X-ray source (SSS) such as V Sge.

  When $\left |\dot M_{\mathrm{acc}} \right |$ is smaller than $\dot
  M_{\mathrm{cr, WD}}$, the following assumptions have been adopted:
      \begin{enumerate}

      \item If $\frac{1}{2}\dot{M}_{\mathrm{cr, WD}} \leqslant
        |\dot{M_{\mathrm{acc}}}| \leqslant \dot{M}_{\mathrm{cr, WD}}$, it is
        assumed that there is no mass loss and that hydrogen-shell
        burning is steady. In this case, before the supernova
        explosion, the system may be observed as the persistent
        SSS \citep{Hach08, Meng10, Kato11}.
           \item If $\frac{1}{8}\dot M_{\mathrm{cr, WD}} \leqslant \left
               |\dot M_{\mathrm{acc}} \right | < \frac{1}{2}\dot
             M_{\mathrm{cr, WD}}$, hydrogen-shell burning is unstable,
             triggering very weak shell flashes, where we assume that
             the processed mass can be retained. Before the supernova
             explosion, this case may be observed as recurrent nova of U
             Sco-type  \citep{Hach08, Meng10, Kato11}.
           \item If $\left |\dot M_{\mathrm{acc}} \right | <
             \frac{1}{8}\dot M_{\mathrm{cr, WD}}$, hydrogen-shell flashes
             are so strong that no mass can be accumulated by the WD.
    
     \end{enumerate}

     These three cases are accounted for in constructing the setup MS
     companion stars for our impact simulations. 
     Furthermore, the mass-growth rate of the WD star was linearly
     interpolated from a grid computed by \citet{Kato04}. The input
     physics in our binary evolution calculations is consistent with
     WLH10 (see also \citealt{Han04}). From our one-dimensional binary
     evolution calculations, we selected ten progenitor systems with a
     representative range of orbital periods and initial companion
     star masses. All resulting models are summarized in Table~\ref{table:1}.

For our hydrodynamical simulation of the impact of SN~Ia ejecta on
their companion stars, we use the GADGET3 code \citep{Spri01,
  Spri05}. Originally, the GADGET code was intended for cosmological
simulations, but it has been modified to make it applicable to stellar
astrophysics problems \citep{Pakm12a}. By using the initial parameters
of the HCV \footnote{In the HCV scenario, a CO WD accretes hydrogen by RLOF from
a lower mass MS secondary. Such a system is formed when a CO WD is left 
in a close binary orbit by an earlier episode of common envelope
evolution in its asymptotic giant branch phase (see \citealt{Mari00}).} scenario of M00, PRWH08 showed that the
SPH-based approach is capable of reproducing previous results obtained
with a grid-based 2D scheme by M00.  This confirms
that the SPH approach with the GADGET code captures the
main dynamical effects of the supernova impact on its companion star.

\subsection{Basic setup}
   \label{sec:setup}

   In our simulation, we use the same method as PRWH08 to map the
   one-dimensional profiles of density, internal energy, and nuclear
   composition of the companion star as obtained from a binary
   evolution calculation to a particle distribution suitable for the
   SPH code. Here, the smoothing length is chosen such that a sphere
   of its radius enclosed 60 neighboring particles.  The rest of the
   basic setup corresponds to that of PRWH08.

   To reduce numerical noise introduced by the mapping to ensure that they are in
   hydrostatic equilibrium before we start the actual simulation, the
   MS companion stars are relaxed for $1.0 \times 10^4\,\mathrm{s}$ (several
   dynamical timescales). If the relaxation succeeds, the
   velocities of the particles stay close to zero. Otherwise, we reject
   the SPH model, and redo the relaxation after adjusting the
   relaxation parameters \citep{Pakm12a}.

   The supernova explosion is represented by the W7 model of
   \citet{Nomo84}. This model has
   been shown to provide a good fit to the observational light curves
   of SNe~Ia. Its total explosion energy is 1.23
   $\times$ 10$^{51}\,\mathrm{erg}$, and the average velocity of the ejecta is $10^4\,\mathrm{km\,s}^{-1}$. We
   place the supernova at a distance to the companion star given
   by the last separation of the binary system. The impact of the
   SN~Ia ejecta on their binary companions is then simulated for
   $5000\,\mathrm{s}$, at which point the mass stripped off from the companion
   star and its kick velocity due to the impact have reached constant
   values (see Section~\ref{sec:test}).

In order to check the effect of the gravitational field of the WD, we run
the impact simulation for the ms\_22a model, including a  $1.4\,M_{\odot}$ WD
during relaxation of the companion star. The ms\_22a model has the smallest
separation (see Table~\ref{table:1}) and should therefore be influenced most strongly by the WD. The
results show that the companion star is basically not distorted due to the
tidal force. The distortion should only be at the percent level in radius anyway,
and our spatial resolution in the very outer layers of the star is not sufficient to
resolve this. However, the mass in these outer layers
is orders of magnitudes smaller than the total stripped mass. Therefore, we run 
all other simulations in this work ignoring the effect of gravitational field of the WD when
the companion stars are relaxed.

\begin{table}
\caption{Resolution test for ms\_22a model.}
\label{table:2}
\centering
\begin{tabular}{c c c c l }     
\hline\hline
  $N_{\mathrm{star}}$ &$N_{\mathrm{total}}$  & $m_{\mathrm{particle}}$ & $M_{\mathrm{stripped}}$ & $v_{\mathrm{kick}}$\\
 & &[$M_{\mathrm{\odot}}$]&[$M_{\mathrm{\odot}}$] &[$\mathrm{km\ s^{-1}}$]\\
\hline
   \ \ \ \ 50\ 000 & \ \ \ \  107\ 092 & 2.41$\times 10^{-5}$ & 0.329 & 138.17\\
   \ \ \ 100\ 000  & \ \ \ 214\ 170    & 1.21$\times 10^{-5}$ & 0.215 & 123.92\\
 \  1\ 000\ 000    & \ 2\ 141\ 336     & 1.21$\times 10^{-6}$ & 0.190 & 111.75\\
 \  2\ 000\ 000    & \ 4\ 282\ 671     & 6.04$\times 10^{-7}$ & 0.185 & 108.90\\
 \  4\ 000\ 000    & \ 8\ 565\ 284     & 3.02$\times 10^{-7}$ & 0.175 & 106.54\\
 \  6\ 000\ 000    & 12\ 847\ 824      & 2.01$\times 10^{-7}$ & 0.173 & 105.29\\
  10\ 000\ 000     & 21\ 413\ 009      & 1.21$\times 10^{-7}$ & 0.173 & 105.31\\
\hline
\end{tabular}

\tablefoot{ $N_{\mathrm{star}}$ and $N_{\mathrm{total}}$ are the number of particles 
  used to represent the companion star and the binary system, respectively.
 All particles have
 the same mass $m_{\mathrm{particle}}$. 
}
         
\end{table}

\section{Simulations}
 \label{sec:numerical_results}

 In this section, we present the numerical results of our SPH impact
 simulations for updated MS companion star models. To ensure the
 reliability of the results from our simulations, we also perform a
 numerical convergence test for one selected companion star model. We
 then explore the dependence of stripped mass and kick velocity on
 the ratios of initial binary orbital separation to companion radius,
 $a_{\mathrm{f}}/R_2$, for a given companion star.

  \subsection{Typical evolution}

   \label{sec:evolution}

   We discuss the evolution of the model ms\_22a
   (Table~\ref{table:1}) as a typical case.
   Figure~\ref{Fig:dens_ms22a} shows the snapshots from our SPH impact
   simulation.  We use a total of $6 \times 10^6$ SPH particles to
   represent the companion star, which corresponds to $\sim 13$\,million
   total SPH particles being used in the simulation. The supernova is
   represented by the W7 model and set up with an initial separation
   of $1.77 \times 10^{11}\,\mathrm{cm}$ as obtained from the binary evolution simulation.

\begin{table}
\caption{Fitting parameters of equation~(\ref{equation:2}) and (\ref{equation:3}).}
\label{table:3}
\centering
\begin{tabular}{c c c c l }     
\hline\hline
 Models & $a_0$ & $\nu$ & $a_1$  & \ \ $\mu$ \\
\hline
   ms\_22a  & 4.092 & -3.137 &  3.84$\times 10^{7}$ & -1.309\\
   ms\_24a  & 3.786 & -3.156 &  4.11$\times 10^{7}$ & -1.509\\
   ms\_28a  & 2.503 & -3.095 &  3.38$\times 10^{7}$ & -1.573\\
   rp3\_20a & 6.105 & -3.489 &  6.05$\times 10^{7}$ & -1.450\\

\hline
\end{tabular}
\end{table}

 \begin{figure}
   \centering
   \includegraphics[width=0.5\textwidth]{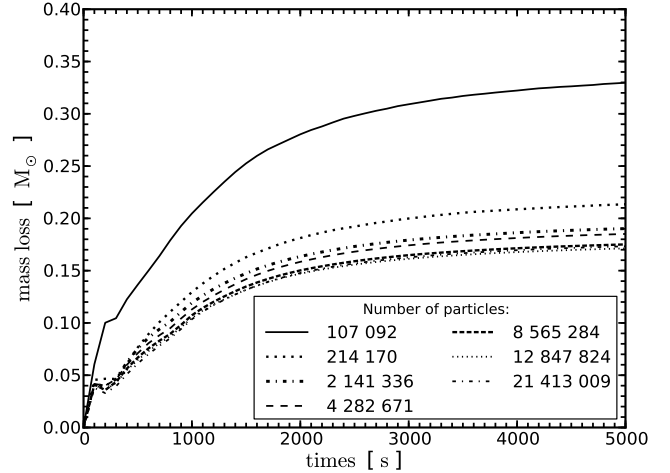}
   \caption{Temporal evolution of the mass loss from the companion
     star for simulations with different numbers of SPH particles.
     Note that the number of particles gives the total of particles in the simulation (both supernova and companion star).  }
         \label{Fig:res_ms22a}
   \end{figure}

 \begin{figure}
   \centering
   \includegraphics[width=0.5\textwidth]{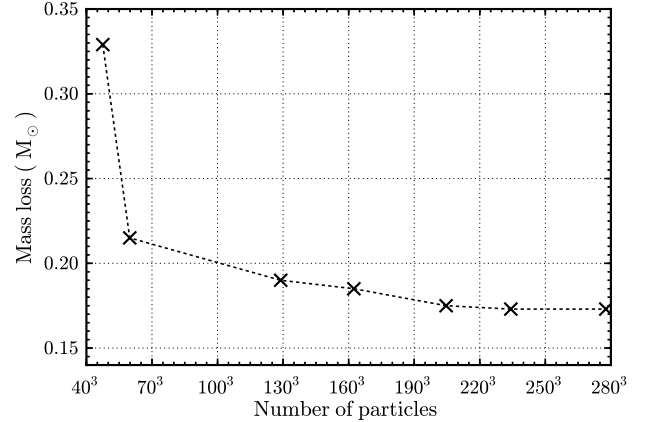}
   \caption{Mass loss vs. different number of SPH particles used in ms\_22a model simulations.}
         \label{Fig:res_mass}
   \end{figure}

  \begin{figure*}
   \centering
   \includegraphics[width=0.98\textwidth]{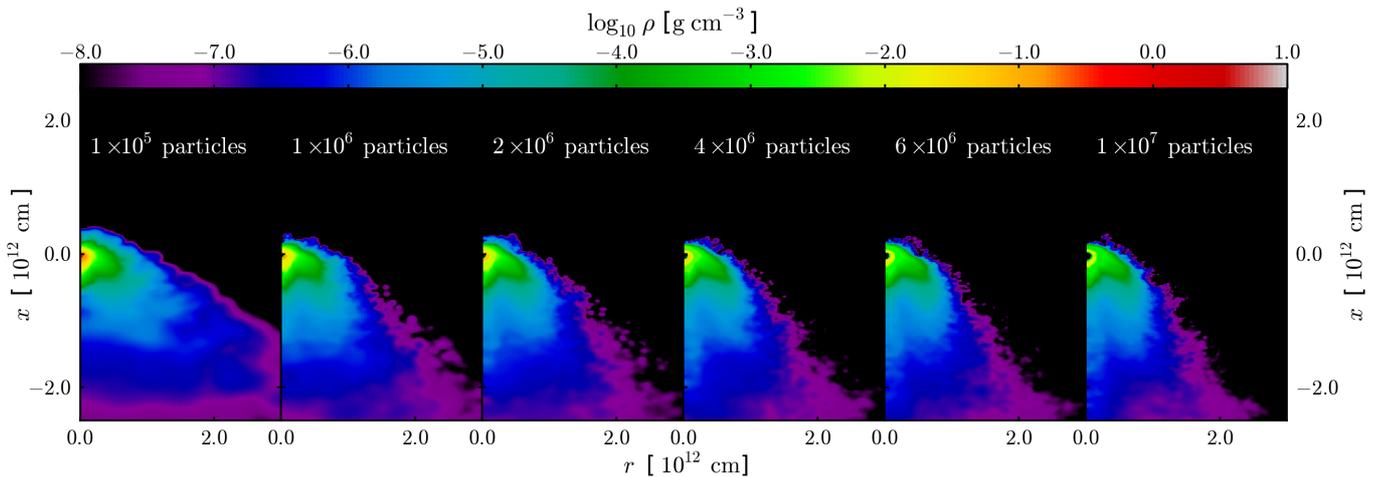}
   \caption{Density distribution of the companion star in the ms\_22a
     model at $3000\,\mathrm{s}$ after supernova explosion with different
     resolutions. Density is color-coded logarithmically. }
         \label{Fig:dens_res_ms22a}
   \end{figure*}

   \begin{figure}
   \centering
   \includegraphics[width=8cm]{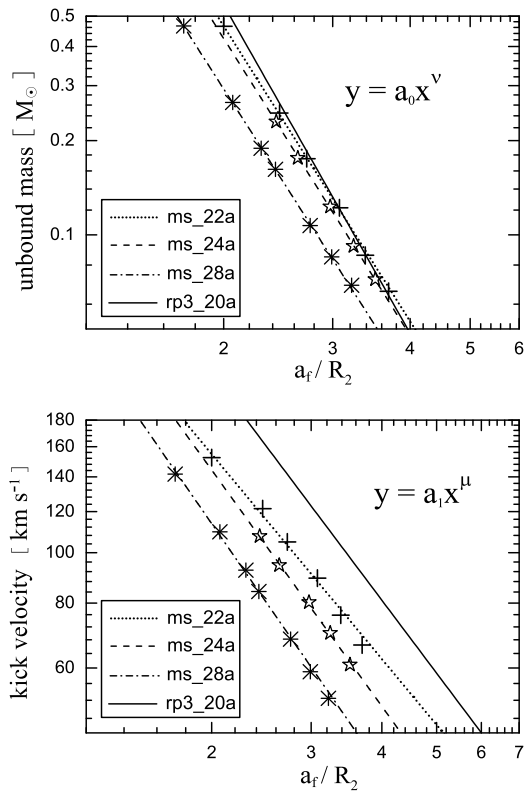}
   \caption{Final stripped mass and kick velocity vs. the ratio of
     orbital separation to the companion radius for different
     companion models. The data are fitted by using the power law of
     equation~(\ref{equation:2}) or (\ref{equation:3}) in this paper.
     The solid line corresponds to the  rp3\_20a model of  PRWH08. Note that
     we use logarithmic coordinates here. All fitting parameters
     are given in Table~\ref{table:3}.  }
         \label{Fig:fit}
   \end{figure}

 \begin{figure}
   \centering
   \includegraphics[width=9.5cm]{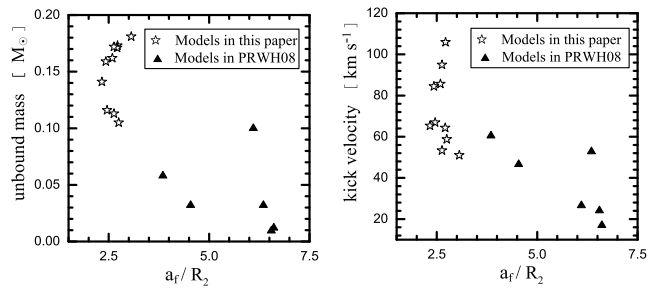}
   \caption{Final stripped mass and kick velocity versus the ratio of
     separation to the radius of companion star,
     $a_{\mathrm{f}}/R_{2}$, for different companion star models. The
     pentacles and filled triangles denote the models simulated in
     this paper and models in PRWH08, respectively.  }
         \label{Fig:a_r}
   \end{figure}

Figure~\ref{Fig:dens_ms22a} illustrates the density distribution of
the companion star for model ms\_22a from the onset of the supernova
explosion until the star starts to relax again ($5000\,\mathrm{s}$).  At the 
beginning of the impact simulation, the MS star is in equilibrium. 
The supernova explodes on the right side of the companion
star. After the explosion, the SN ejecta expand freely, approach the
companion star from the right side, and a shock wave develops as they
hit its surface (see first snapshot). The second
snapshot shows the impact $100\,\mathrm{s}$ after the supernova explosion, when
the shock starts to propagate
through the companion star. At $400\,\mathrm{s}$ (third snapshot), the shock wave reaches its center. The supernova
ejecta flow around the companion and a hole forms in them.  In the
fourth snapshot ($1500\,\mathrm{s}$), the shock wave in the companion has
passed the stellar core, and the supernova ejecta are
mixed with hydrogen-rich material stripped from the companion
star. The last two snapshots show the interaction at $3000\,\mathrm{s}$ and $5000\,\mathrm{s}$
after SN explosion. This is the end of the phase of mass-stripping by
the impact; the remnant of the companion star shrinks and relaxes to
be almost spherical again. The mass stripped by the impact of the SN
ejecta stays constant from this time onwards (see
Fig.~\ref{Fig:res_ms22a}).  In
Fig.~\ref{Fig:dens_ms22a}, it can be seen that the companion star has
moved to left by $\sim 4.5\times 10^{10}\,\mathrm{cm}$ from $0.0\,\mathrm{s}$ to $5000\,\mathrm{s}$
due to the kick caused by the impact.

  \subsection{Numerical convergence test}
    \label{sec:test}

    To ensure reliability of our numerical results, we
    perform a convergence test.  We use model ms\_22a and carry out the
    simulations for different resolutions ranging
    from $1.07 \times 10^{5}$ to $2.14 \times 10^7$ SPH particles
    (see Table~\ref{table:2}).  For each resolution, the mass stripped
    from the companion star at different times after the supernova explosion
    is calculated (see Fig.~\ref{Fig:res_ms22a}).  The total unbound mass and the kick
    velocity obtained from the impact $5000\,\mathrm{s}$ after explosion with
    different resolutions are listed in Table~\ref{table:2}. The
    stripped mass is calculated by summing up the mass of all
    particles that are not bound to the star any more, but were part
    of the star in the initial setup. In order to determine whether
    or not a particle is bound to the star, we calculated the sum of
    the kinetic energy (positive) and potential energy (negative) for
    each particle. If the total energy is positive, the particle is
    not bound.  Note that the center-of-mass motion of star is
    subtracted when calculating the kinetic energy for each particle.

   Figure~\ref{Fig:res_ms22a} shows that the mass loss decreases when 
   the number of the SPH particles used in the simulations
   increases (see also Fig.~\ref{Fig:res_mass}). However, it 
   is numerically well converged for more than about 8 million SPH particles in the
   simulation (this corresponds about 4 million particles in the companion 
   star). The difference in stripped mass between simulations with $8.56 \times 10^{6}$ (the short dashed line) 
   and $1.28 \times 10^7$ (the dotted line)
   SPH particles is smaller than $2\%$, the difference between $1.28 \times 10^6$ and $2.14 \times 10^7$ (the narrow dash-dotted line)
   SPH particles is less than $1\%$ (see Table~\ref{table:2}).  After $3000\,\mathrm{s}$, the amount of
   unbound mass becomes constant in good approximation. The density
   distribution of the companion star corresponding to different
   resolutions is shown in Fig.~\ref{Fig:dens_res_ms22a}.  There
   are also no morphological differences among the simulations with 
   $8.56 \times 10^6$ , $1.28 \times 10^7$  and $2.14 \times 10^7$ SPH particles (the last
   three snapshots in Fig.~\ref{Fig:dens_res_ms22a}). Therefore, we
   conclude that it is sufficient to represent the companion stars with about 6 million
   SPH particles in our SPH impact
   simulations.

  \subsection{Results}
  \label{sec:results}

  Based on a selection of ten realistic MS companion star models we
  simulate the interaction of supernova ejecta with their binary
  companions. The initial parameters of all binary systems used in
  this paper are listed in Table~\ref{table:1}.

  The amount of hydrogen-rich material stripped from the surface of
  the companion stars, $M_{\mathrm{stripped}}$, and the kick velocities caused
  by the supernova impact, $v_{\mathrm{kick}}$, are also shown in
  Table~\ref{table:1} (see also Fig.~\ref{Fig:fit}). The stripped masses range from $0.11$ to
  $0.18\,M_{\mathrm{\odot}}$ and we measure kick velocities between
  $51$ and $105\,\mathrm{km\,s^{-1}}$. Note that this kick velocity is
  defined as the center of mass velocity of all particles bound to the
  companion star $5000\,\mathrm{s}$ after the explosion of the
  supernova.

\begin{table}
\caption{Stripped mass.} \label{table:4}
\centering
\begin{tabular}{c c c  c c c}     
\hline\hline
                
Model & $M_{\mathrm{c,f}}$  & $a_{\mathrm{f}}/R_2$  &\ $\Delta M^{\mathrm{1}}$ & $\Delta M^{\mathrm{2}}$ & Difference\\
  &[$M_{\mathrm{\odot}}$] &  & [$M_{\mathrm{\odot}}$]& [$M_{\mathrm{\odot}}$] & [\%] \\
\hline
   ms\_20a & 0.74 & 3.06 &\  \ 0.18 & 0.19 & 5\\
   ms\_20b & 1.17 & 2.75 &\  \ 0.11 & 0.14 & 27\\
   ms\_22a & 1.21 & 2.72 &\  \ 0.17 & 0.18 & 6\\
   ms\_20c & 1.22 & 2.71 &\  \ 0.17 & 0.14 & 18\\
   ms\_24a & 1.40 & 2.63 &\  \ 0.17 & 0.19 & 12\\
   ms\_20d & 1.40 & 2.63 &\  \ 0.11 & 0.15 & 36\\
   ms\_23a & 1.50 & 2.59 &\  \ 0.16 & 0.19 & 19\\
   ms\_24b & 1.88 & 2.46 &\  \ 0.12 & 0.17 & 42\\
   ms\_28a & 2.00 & 2.43 &\  \ 0.16 & 0.21 & 32\\
   ms\_30a & 2.45 & 2.33 &\  \ 0.14 & 0.20 & 43\\
\hline
\end{tabular}
\begin{list}{}{}
\item[$^{\mathrm{1}}$] Numerical calculation of stripped mass in our simulations.
\item[$^{\mathrm{2}}$] Stripped mass analytically estimated by using the method of 
                   \citet{Whee75}. Here, we directly use the density profiles of companion
                   models from the 1D stellar evolution calculations.
\end{list}
\end{table}

  The largest stripped mass of our simulations is found for model
  ms\_20a ($0.18\,M_\odot$).  Further analysis shows that the
  hydrogen in the center of this companion star has already been
  mostly depleted at the time the white dwarf explodes in a supernova,
  and its outer layers already begin to expand. This moves the star on
  its evolutionary track in the Hertzsprung-Russell (H-R) diagram
  towards the giant phase. The density profile of companion star
  ms\_20a is shown in next section. Compared with other main
  sequence stars in our sample, this star has a very large
  radius and a higher density in the core. The shrinking
  of its inner core and the subsequent expansion of the outer layers
  make its envelope less bound. This explains why more hydrogen-rich
  material is stripped off when the supernova blast wave hits the
  companion star in this model.

   \begin{figure}
   \centering
   \includegraphics[width=8cm]{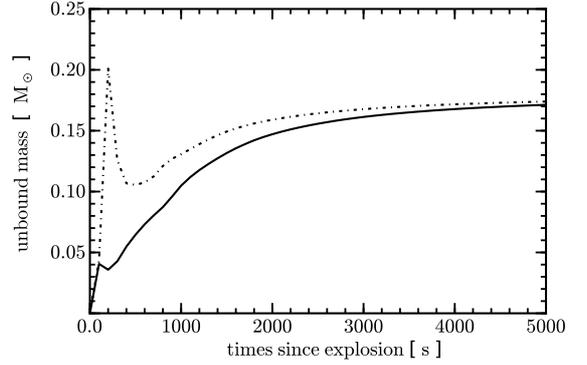}
   \caption{Unbound mass vs.\ simulation time in ms\_22a model. The
     solid line shows the total mass of all particles with a total
     energy (kinetic plus potential energy) larger than zero. The
     dash-dotted line also includes the internal energy in the sum. }
              \label{Fig:unboundmass}%
    \end{figure}

 \begin{figure}
   \centering
   \includegraphics[width=9cm]{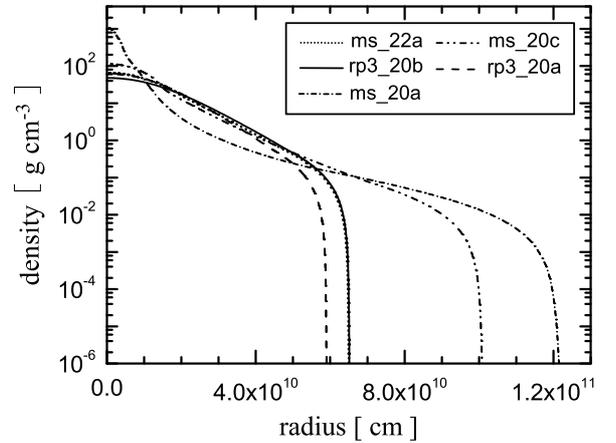}
   \caption{Initial density profiles of five companion star models at
     the onset of the SN~Ia explosion.  Model rp3\_20a (dashed line) and
     rp3\_20b (solid line) are from PRWH08. }
         \label{Fig:dens_r}
   \end{figure}

 \begin{figure*}
   \centering
   \includegraphics[width=14cm]{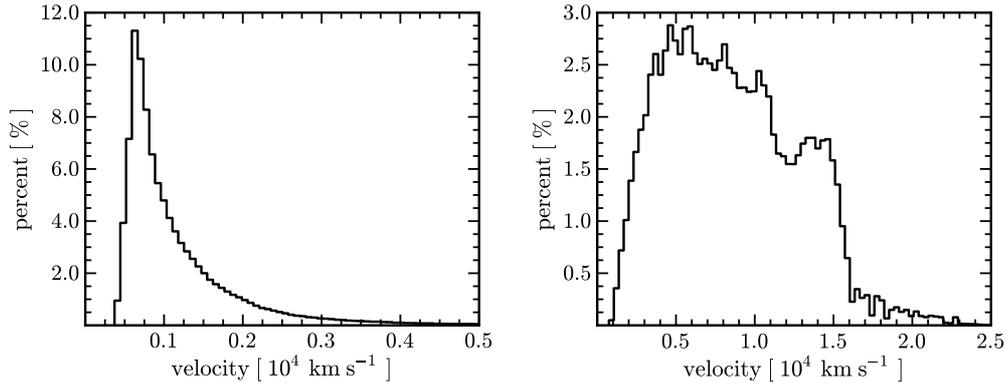}
   \caption{Final velocity distribution of the stripped hydrogen-rich
     material that originally belonged to the companion star (left
     panel) and supernova ejecta (right panel) $5000\,\mathrm{s}$
     after the supernova explosions for model ms\_22a.  }
         \label{Fig:vel_ms22a}
   \end{figure*}

  For a given companion star model, we investigate the dependence of
  stripped mass, $M_{\mathrm{stripped}}$, and kick velocity,
  $v_{\mathrm{kick}}$, on the ratio of orbital separation to companion
  star radius, $a_{\mathrm{f}}/R_2$.  All parameters but the orbital
  separation are kept constant (it means that we only change 
  the value of the separation, $a_{\mathrm{f}}$, for the same companion model artificially). 
  Figure~\ref{Fig:fit} shows the
  stripped mass and kick velocity versus $a_{\mathrm{f}}/R_2$ for models ms\_22a,
  ms\_24a and ms\_28a. For a given companion star, we
  find a similar relationship as previous studies \citep{Mari00,
    Meng07, Pakm08, Pan12}. The dependence of the stripped mass on this ratio
  follows a power law to good approximation (see Fig.~\ref{Fig:fit}):

   \begin{equation}
    \label{equation:2}
     M_{\mathrm{stripped}}= a_0 \cdot \ \left(\frac{a_{\mathrm{f}}}{R_2}\right)^{\nu} \ \ M_{\mathrm{\odot}}.
   \end{equation}
   Likewise, the dependence of the kick
   velocity, $v_{\mathrm{kick}}$ on $a_{\mathrm{f}}/R_2$ can be fitted by a power law (see
   Fig.~\ref{Fig:fit})

  \begin{equation}
    \label{equation:3}
     v_{\mathrm{kick}}= a_1 \cdot \ \left(\frac{a_{\mathrm{f}}}{R_2}\right)^{\mu} \ \ \mathrm{cm\ s^{-1}}.
   \end{equation}
   Here, $a_0$ and $a_1$ are two fitting constants, which, however,
   are not unique but depend on the companion star models.  The 
   parameters $\nu$ and $\mu$ are the corresponding power law indices. 
   Values for these fitting parameters are listed in Table~\ref{table:3}. For
   comparison, we also show the result from model rp3\_20a of PRWH08.
   Comparing our models with PRWH08, all simulations follow a power law
   of some kind, but we do not obtain the same fitting parameters. This implies 
   the importance of the structure of the companion star for the 
    stripped mass and kick velocity in
   our impact simulations. 
   Furthermore, we put all models used in this paper and the PRWH08 models together as a whole sample
   to examine the effect of $a_{\mathrm{f}}/R_{2}$. Figure~\ref{Fig:a_r} shows the dependence of the stripped mass
   and kick velocity on the parameter, $a_{\mathrm{f}}/R_{2}$, for all these models. 
   However, we do not find a  power law relation is still holding in Fig.~\ref{Fig:a_r}. 
   Therefore, again, it indicates the results of our impact simulations are also dependent on
   the details of the structure of the companion
   stars due to the history of mass transfer.


\section{Discussion}
  \label{sec:discussion}

 \subsection{Comparison with previous studies}
  \label{sec:comparison}

  \citet{Whee75} derived a simple analytic formula for the calculation of stripped mass, 
  and their results were confirmed by some of the early numerical simulations (see \citealt{Mari00}).
  Here, we also estimate the stripped mass using their analytic method based on our 
  MS companion models (see Table~\ref{table:4}). As shown in Table~\ref{table:4}, 
  the order of magnitude of stripped mass predicted by the analytic estimate agrees with our results, 
  but is usually overestimated by the analytic 
  formula compared to the result obtained from our simulations, with deviations ranging from $5\%$ to $43\%$.

  Using a $1.0\,M_{\mathrm{\odot}}$ solar-like companion star model,
  \citet{Mari00} found about $0.15\,\mathrm{M}_{\mathrm{\odot}}$
  hydrogen-rich material to be stripped from the surface of MS
  companion by the impact of the supernova explosions. The
  mass of stripped hydrogen-rich material in our simulation is
  consistent with their results.  Note that compared to the study of PRWH08
  our simulations bring the
  results back to the original work of \citet{Mari00} only by chance
  -- the setups of the companion stars are very different. While we construct our companion
  star models from consistent binary evolution calculations, their model ignores the effect of
  mass transfer on the structure of the companion
  star altogether. Although PRWH08 included such an effect in the
  construction of their companion stars, it was not done in a
  consistent binary evolution calculation.

  The amount of the unbound mass in our work is also
  consistent with the new multi-dimensional adaptive mesh refinement
  simulations of \citet{Pan12}.
  But, again, they did not follow the full binary evolution but used initial
  conditions with a constant mass-loss rate when constructing
  their MS companion star models with the MESA code of \citet{Paxt11}.
  Therefore, the agreement can be regarded as coincidental, too. \citet{Pan12} argue that
  the mass stripping is dominated by ablation in their simulations for
  the MS companion. They used a slightly different criterion for the unbound
  mass in which they included the internal energy in addition to the potential and
  kinetic energy. This does not make an effect for our simulations because
  we are mainly interested in the unbound mass at late enough times when the internal energy is
  negligible to the kinetic energy.
  Figure~\ref{Fig:unboundmass} shows the unbound mass of a companion
  star as a function of  time after the supernova explosion in our SPH
  simulation. Whether or not a particle is unbound is decided by
  summing its potential energy, kinetic energy and internal energy (or
  without the internal energy, which corresponds to the solid line in
  Fig.~\ref{Fig:unboundmass}).  Already
  $5000\,\mathrm{s}$ after the explosion most of the internal energy
  deposited by the impact has been converted into kinetic
  energy. Therefore, as in PRWH08, we neglect the internal energy when
  we flag particles as bound or unbound at late times in our
  simulation.

  In our study we find that a minimum of $0.1\,\mathrm{M}_{\odot}$ of
  hydrogen-rich material being stripped from the companion star.  This
  is significantly larger than the $0.035\,\mathrm{M}_{\mathrm{\odot}}$
  for stripped hydrogen found by \citet{Meng07}. This might be caused
  by their oversimplified description of the interaction physics, e.g.\ the 
  effect of the shock formed between the supernova ejecta and the companion 
  star was not calculated in their analytic model. Recently, \citet{Pan12} argued that 
  since they neglected the mass loss due to the ablation from the 
  hot surface of the companion star, \citet{Meng07} underestimate the
  final unbound mass.

   PRWH08 found stripped masses in the range from
   $0.01\,\mathrm{M}_{\mathrm{\odot}}$ to $0.06\,\mathrm{M}_{\mathrm{\odot}}$,
   which is very close the observational limit obtained by
   \citet{Leon07}, but significantly lower than the values we find for
   the models presented here. In order to determine the origin of this
   difference, we select two models, ms\_20c and ms\_22a (see
   Table~\ref{table:1}), in comparison to the models rp3\_20a and rp3\_20b of PRWH08.
   The density profiles of these four models at
   the moment of the explosion of the supernova are shown in
   Fig.~\ref{Fig:dens_r}. Models ms\_22a and rp3\_20b have the same radius of $0.65 \times
   10^{11}\,\mathrm{cm}$. Although there are some small differences in the
   density profiles in the inner cores, the outer
   layers are nearly identical. In our simulations, only the
   properties of outer layers of the companions can significantly
   affect the results of the
   interaction between supernova ejecta and companion
   star. Therefore, we chose these two models to carry out a
   comparison. PRWH08 had set up the binary
   with a separation of $4.26 \times 10^{11}\,\mathrm{cm}$ for their Model
   rp3\_20b. They found a stripped mass of
   $0.01\,\mathrm{M}_{\mathrm{\odot}}$ and a kick velocity of $24.1\,\mathrm{km\,s^{-1}}$. 
   Our model ms\_22a, however, has an orbital separation at
   the time of the explosion of $1.77 \times 10^{11}\,\mathrm{cm}$ only.  Running
   this model we obtain a significantly larger stripped mass of $0.17\,M_{\mathrm{\odot}}$ 
   and a higher kick velocity of $105.29\,\mathrm{km\,s^{-1}}$.  This is to be expected for changing
   the initial separation, as discussed above.  Furthermore, we
   calculate the stripped mass and kick velocity with the same
   separation $4.26 \times 10^{11}\,\mathrm{cm}$ as for Model rp3\_20b by using
    the power law relation of equation~(\ref{equation:2}) and
   (\ref{equation:3}) for our Model ms\_22a. We find a stripped mass
   of $0.01\,M_{\mathrm{\odot}}$ and kick velocity of $33.02\,\mathrm{km\,s^{-1}}$,
   which is excellent agreement with Model rp3\_20b. Therefore,
   excluding the density effect, the orbital separation at the time of
   the explosion is the primary factor to cause the difference between
   the simulations of the two models ms\_22a and rp3\_20b.

   The difference in orbital separations originates from different
   treatments of the progenitor evolution. The binary systems we 
   examine here fill their Roche-lobe at the time of the explosion. This 
   fixes the orbital separations of the binary systems. 
   Instead of detailed binary evolution calculations, PRWH08 directly
   took parameters from the study of \citet{Ivan04} to mimic the
   effect of the binary evolution phase. \citet{Ivan04} analyzed the
   evolution of possible SN~Ia progenitor systems consisting of a WD
   and an evolved MS star. In PRWH08, all values adopted by model
   rp3\_20a come from a WD+MS binary system evolved by
   \citet{Ivan04} with initial WD mass $M_{\mathrm{WD}}$ = $0.8\,M_{\odot}$,
   companion mass $M_{\mathrm{d,i}}$ = $2.0\,M_{\odot}$ and orbital period
   $P_{\mathrm{i}}$ = $1\,d$ \citep{Ivan04}.  These parameters have been
   presented in Table~2 of PRWH08. For comparison,
   we set up our new Model ms\_20c with the same initial binary
   parameters ($M_{\mathrm{WD}}$ = $0.8\,M_{\odot}$, $M_{\mathrm{d,i}}$ = $2.0\,M_{\odot}$ 
   and $P_{\mathrm{i}} = 1\,d$ ) and carry out the
   fully detailed binary evolution calculation to construct the
   structure of companion star. Properties of Model
   ms\_20c are listed in Table~\ref{table:1}. The final orbital
   separation at the moment of the supernova explosion is $2.74 \times 10^{11}\mathrm{cm}$, 
   which is very close to the $2.68 \times 10^{11}\,\mathrm{cm}$ 
    of Model rp3\_20a by PRWH08. Moreover, the final
   companion masses agree very well ($1.22\,M_{\odot}$ for our Model ms\_20c versus $1.17\,M_{\odot}$
   for Model rp3\_20a).
   However, Fig.~\ref{Fig:dens_r} shows that the ms\_20c model
   has a larger radius compared to Model rp3\_20a. Therefore, it
   is not surprising that the impact simulation for this model leads
   to a stripped mass of $0.171\,M_{\mathrm{\odot}}$ (see
   Table~\ref{table:1}), while only $0.032\,M_{\mathrm{\odot}}$
   were stripped in the rp3\_20a model.  Since our Model ms\_20c has a
   larger radius, it has a less bound envelope that can be stripped
   away more easily. Moreover, the larger radius causes an extended
   interaction area that also leads to a slightly larger kick velocity
   of $64.32\,\mathrm{km\,s^{-1}}$ compared with the $46.6\,\mathrm{km\,s^{-1}}$ of Model
   rp3\_20a.  Finally, in our model, a larger conical hole is created in the supernova
   debris behind the companion star.  Without the effect of orbital
   separation, the more compact structure of the companion star
   significantly reduces the stripped mass. Thus, the degree of
   compactness of a companion star, especially the compactness of its
   outer layer is very important to determine the influence of the impact of the SN~Ia
   ejecta on its binary companion star.

 \begin{figure}
   \centering
   \includegraphics[width=0.45\textwidth]{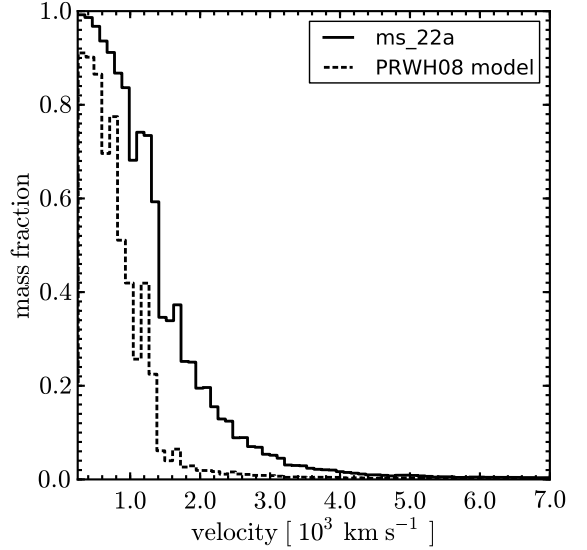}
   \caption{Fraction of the material stripped from the companion in
     velocity space for ms\_22a model $5000\,\mathrm{s}$ after the
     supernova explosion relative to the total contaminated supernova
     ejecta. The dashed line corresponds to the PRWH08 model.}
         \label{Fig:mass_frac_ms22a}
   \end{figure}

 \begin{figure*}
   \centering
   \includegraphics[width=12cm]{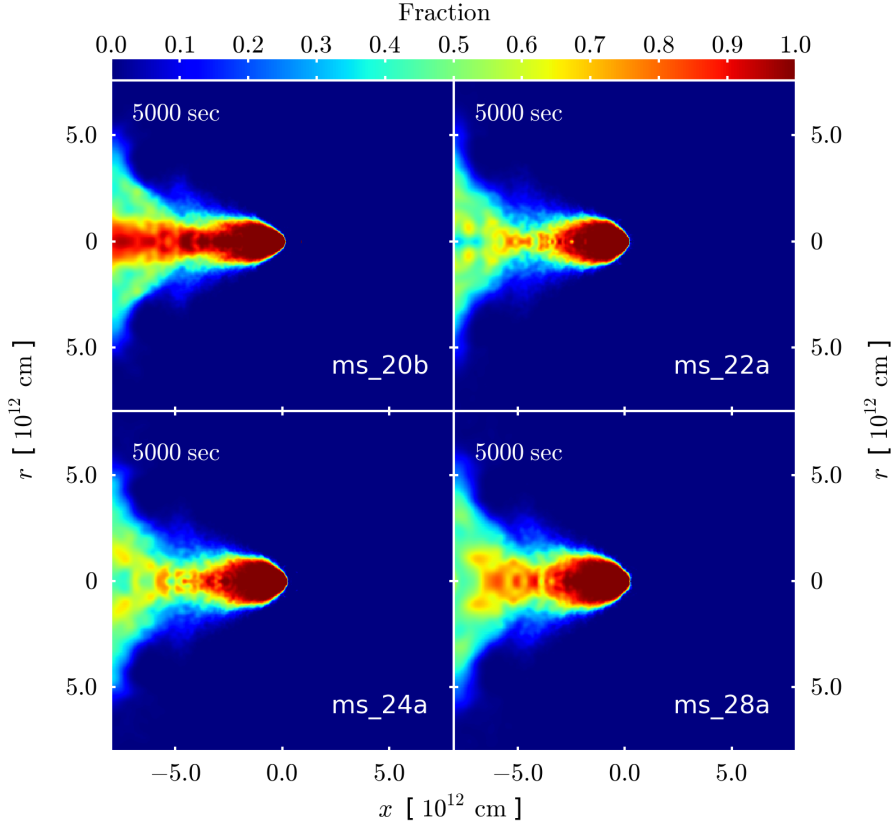}
   \caption{Relative fraction of material that belonged originally to
     the companion with respect to the total amount of material. The
     red and blue color correspond to companion and supernova
     material, respectively.  }
         \label{Fig:mix4}
   \end{figure*}

 \begin{figure*}
       \centering
   \includegraphics[width=12cm]{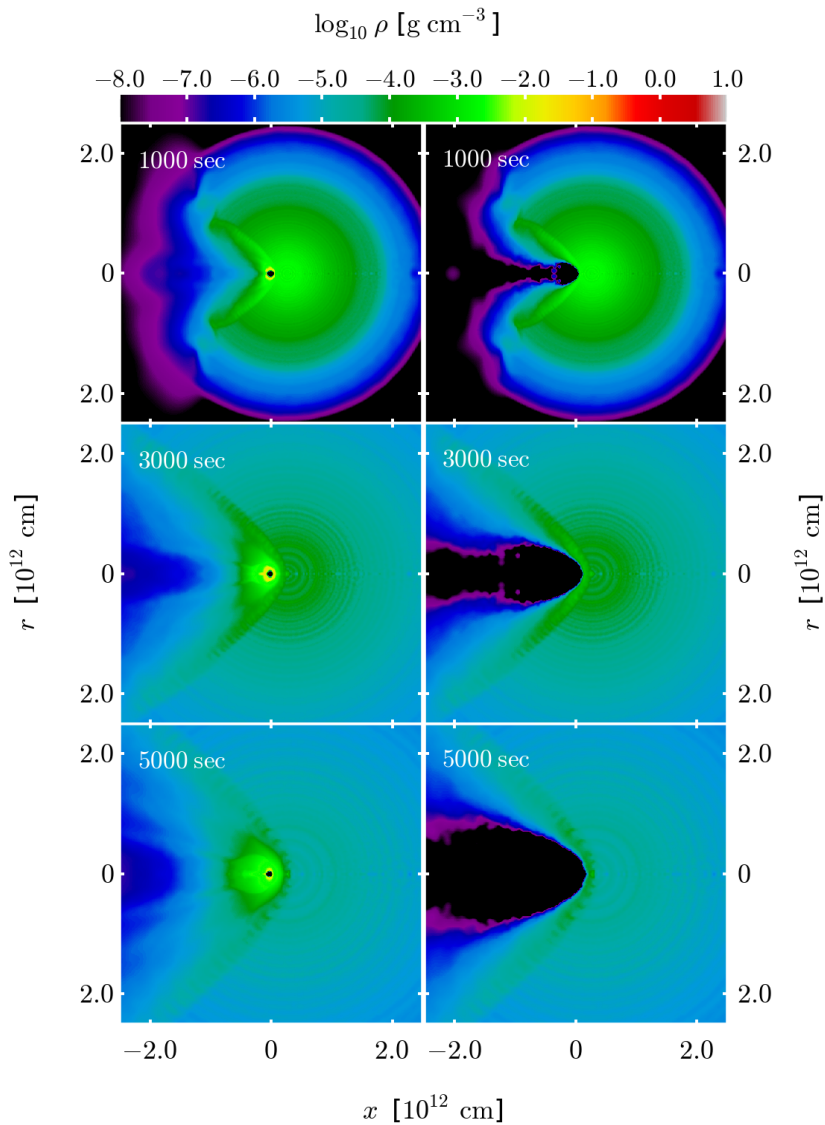}
   \caption{Density distributions of all material (left panel) and
     the supernova ejecta (right panel) at $1000\,\mathrm{s}$, $3000\,\mathrm{s}$ and $5000\,\mathrm{s}$ after
     the explosion for model ms\_20c. The logarithm of density is color-coded. }
         \label{Fig:ejecta_ms20c}
   \end{figure*}

 \begin{figure*}
   \centering
   \includegraphics[width=12cm]{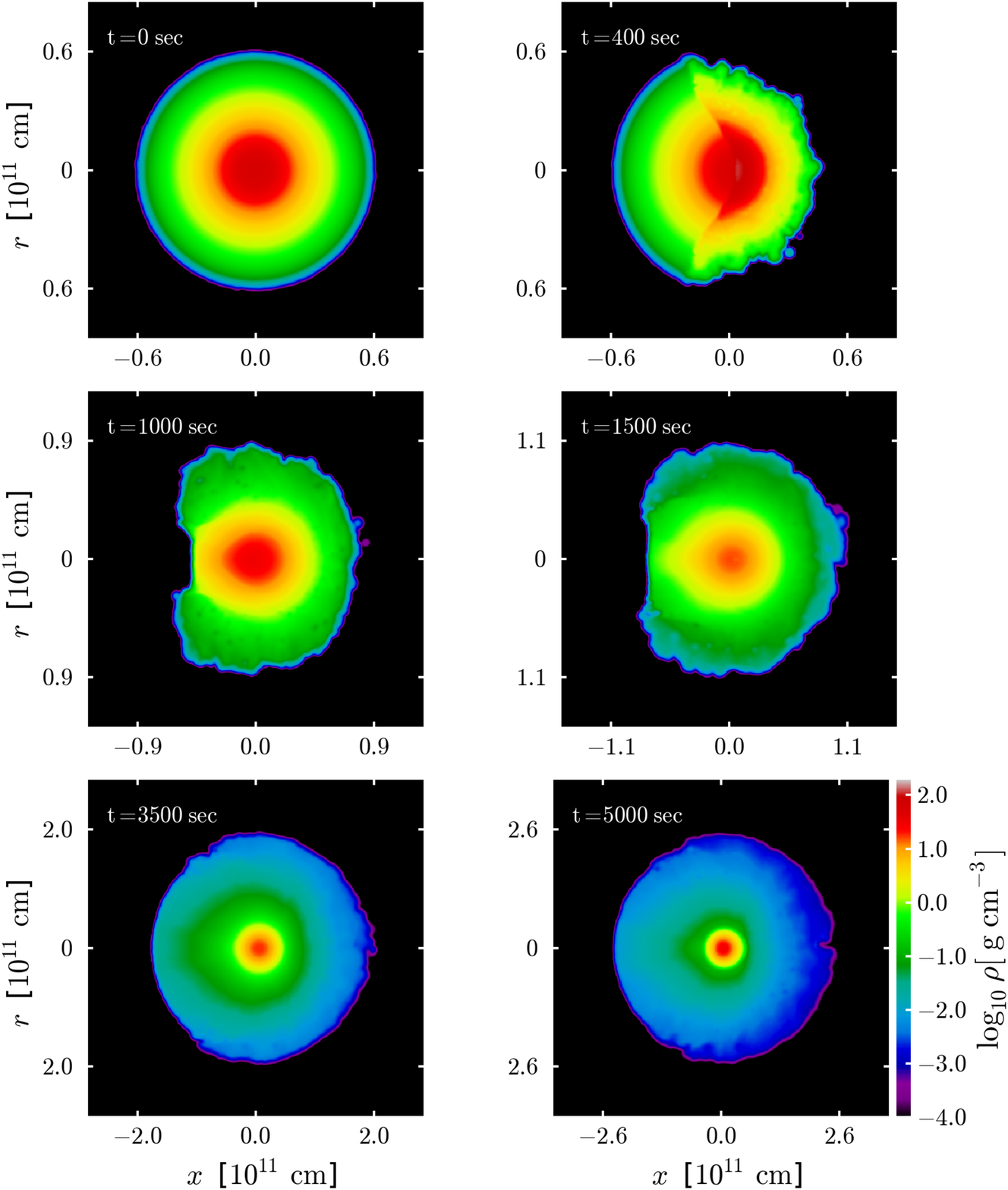}
   \caption{Density evolution of the remnant star in  ms\_22a model simulations. Only the bound material
            that originally belonged to the companion star are included. The logarithm of density is color-coded.}
         \label{Fig:remnant}
   \end{figure*}

   What causes this difference of companion star structures between the models
   ms\_20c and rp3\_20a? PRWH08 computed their companion star
   models by constantly removing mass while they evolved a single
   main sequence star.  They did not consider the detailed
   mass-transfer processes in a binary system. In their
   one-dimensional stellar evolution, the mass loss proceeds rather
   rapidly: the duration of the mass transfer is about a factor of 10
   less than the Kelvin-Helmholtz timescale of their stars.  Although
   they mimicked the total mass loss and the total time of mass loss
   from \citet{Ivan04} when constructing their companion star
   models, this does not lead to realistic companion star structures
   such as obtained from consistent binary evolution. The result are more
   compact objects and the binary systems of PRWH08 are characterized by very large
   values of the parameter, $a_{\mathrm{f}}/R_{\mathrm{2}}$, since they used less evolved companion
   stars to start the mass accretion phase from, which were too compact to actually fill their
   Roche-lobe (see Fig.~\ref{Fig:a_r}).  
   In contrast, RLOF is taken into account
   for the mass transfer of a binary system in our detailed binary
   evolution calculations, and we also consider the process of WD
   accretion by including the optically thick wind model of
   \citet{Hach99b}. The structures of companion stars are
   properly adjusted due to the detailed modeling of mass transfer in the binary
   system. Consequently, our work presents an update to the PRWH08 study with more
   realistic companion star models.

   In our SPH impact simulations, the stripped mass and the kick
   velocity depend sensitively on the ratio of the orbital separation to the
   radius of companion star, $a_{\mathrm{f}}/R_{2}$. More compact
   companion stars with a larger orbital separation lead to a
   significant reduction of the amount of stripped hydrogen-rich
   material. However, $a_{\mathrm{f}}/R_{2}$ is not the
   only parameter to determine the outcome of the supernova
   impact, the companion structure is also important (see Fig.~\ref{Fig:a_r}).

  \subsection{Distribution of the stripped material in velocity space}

    \label{sec:vel}

    The hydrogen-rich material stripped from the companion star is
    mixed into the supernova ejecta. Figure~\ref{Fig:vel_ms22a} shows
    the velocity distribution of the hydrogen-rich material stripped
    from the companion star (left panel) and the supernova ejecta (right
    panel).  Most of the stripped material is concentrated at
    velocities below $\sim 800\,\mathrm{km\,s^{-1}}$, which is much slower than 
    the typical velocities of the supernova ejecta of $\sim 10~000\,\mathrm{km\,s^{-1}}$\citep{Chug86, Mari00},
    placing it at the very center of the ejecta.

    In Fig.~\ref{Fig:mass_frac_ms22a} we plot the mass fraction of
    material stripped from the companion star in velocity space 
    (all bound material has been cut out). We also show the
    result of PRWH08 (dashed line) for comparison. In the low
    velocity region, the mass fraction of the stripped material is
    very high and it dominates over the original supernova
    material. But this fraction sharply decreases as the velocity
    becomes larger than $10^3\,\mathrm{km\,s^{-1}}$. Clearly, the supernova
    ejecta dominate at high velocities. However, some stripped
    material is present also in the outer ejecta as already noted by M00.
    They argued 
    that the presence of stripped material at high velocities implies 
    that traces of hydrogen from the companion are swept up
    in the oxygen and silicon layer of the supernova ejecta. They also
    argued that given the upper limits on the hydrogen abundance from
    SN~Ia observations near maximum light, this may provide a
    criterion for discriminating between SN~Ia progenitor scenarios
    (M00).  Figure~\ref{Fig:mix4} illustrates the relative
    amount of material that originally belonged to the companion with
    respect to the total amount of material for different companion
    star models. This figure shows how the supernova ejecta are mixed with the
    material stripped from the surface of the companion star.  
    Most of the stripped material, however, will
    only become visible at very late times when the ejecta are mostly
    transparent and it is possible to see very deep into the center of
    the ejecta.

    Figure~\ref{Fig:ejecta_ms20c} shows how the spatial distribution
    of stripped material evolves with time for model ms\_20c. The
    left column shows both, the SN~Ia and companion star at $1000\,\mathrm{s}$,
    $3000\,\mathrm{s}$ and $5000\,\mathrm{s}$, respectively. The right column shows the
    material originally belonging to the supernova only --  all companion
    material has been cut out. We see that the supernova ejecta are
    significantly affected by the companion star when the supernova
    impacts the binary companion.  The interaction creates a conical
    hole in the supernova debris with an opening angle of about
    $50^{\circ}$ (see the right snapshot at $1000\,\mathrm{s}$ of
    Fig.~\ref{Fig:ejecta_ms20c}). Comparing with PRWH08, our
    simulation shows a larger cone-like hole behind the companion
    star. This is not surprising, because at the same initial
    separation our companion star has a
    larger radius than the models used in PRWH08.  Additionally, Fig.~\ref{Fig:ejecta_ms20c} shows that
    the material is shocked at the companion star which leads to
    the formation of a bow shock. Ejecta passing through the shock 
    are heated and compressed into a thin shell, and their velocity
    vector is redirected \citep{Kase10}.

 \subsection{Detection of hydrogen}
  \label{sec:detection}

  In our simulations, we find that the supernova impact strips off
  $0.11\,M_\odot$ to $0.18\,M_\odot$ of hydrogen-rich material 
  from the companion star (see Table 1). This is far more than
  the most stringent upper limit of $0.01\,M_\odot$ which 
  \citet{Leon07} derived from the non-detection of H$_\alpha$ 
  emission in late time spectra. Therefore our results might 
  challenge the SD scenario if the systems studied here are 
  representative and the objects, from which the observational 
  constraints were derived, originate from SD progenitors.

  However, it is important to note that the model of \citet{Matt05}
  which was employed by \citet{Leon07} to obtain the upper limit 
  mentioned above is quite simple. Moreover, most of the stripped 
  hydrogen in our models ends up at velocities  below $10^{\mathrm{3}}\,\mathrm{km\,s^{-1}}$ 
  so that it is confined to the innermost part 
  of the explosion ejecta which are usually rich in iron-group 
  elements. Whether or not H$_\alpha$ emission will be detectable 
  under these conditions, is a highly complex question which can 
  only be answered by performing sophisticated radiative transfer 
  simulations on the abundance structure of our explosion models.
  This is beyond the scope of the current paper.

  Note also that there are some possibilities to reduce the 
  amount of stripped hydrogen. In our detailed binary evolution 
  calculations, we adopt the prescription of \citet{Hach99a} for 
  the mass growth of a CO WD by accretion of hydrogen-rich material 
  from its companion. According to this model unprocessed matter 
  is assumed to be lost from the binary system due to an optically 
  thick wind if the mass accretion rate exceeds a critical value of
  $\dot{M}_{\mathrm{cr,WD}}= 5.3 \times
  10^{7}(1.7/X-1)(M_{\mathrm{WD}}-0.4)\,M_\odot\mathrm{yr^{-1}}$ (here, $X$ is 
  the hydrogen mass fraction and $M_{\mathrm{WD}}$ the mass of the 
  accreting WD). \citet{Hach99a, Hach08} proposed that this optically 
  thick wind will strip off parts of the outer layer of the companion 
  star. This ``mass-stripping'' effect is currently neglected in our
  simulations. However, it could reduce the amount of hydrogen
  in the SN ejecta since the mass-stripping reduces the companion size 
  and therefore also the mass lost in the supernova impact.

  Another possibility to reduce the amount of stripped hydrogen 
  arises if the MS companion in the binary has a helium-rich 
  envelope (\citet{Pan10} find the stripped helium mass to be 
  consistent with observational constraints in this case). Such 
  a system is possible if a massive primary undergoes 
  a first RLOF during its red giant phase and the resulting helium 
  star experiences a second RLOF episode during core helium burning 
  \citep{Hach99a, Hach99b}.

  Finally, we note that also the recently proposed ``spin-up and 
  spin-down'' model \citep{Just11, Di11, Hach12} is likely to reduce 
  hydrogen stripping.

\subsection{Surviving companion stars}

In the SD scenario of SNe~Ia, the companion star survives the
supernova explosion.  Because the interaction with the SN~Ia ejecta
and its orbital velocity at the time of the explosion, its spatial
velocity might distinguish it from stars in its neighborhood. In our
simulation, the kick velocity reaches values from $\mathrm{51~km\,s^{-1}}$
to $\mathrm{105~km\,s^{-1}}$, which is comparable to the orbital velocity
of $\mathrm{96-281~km\,s^{-1}}$.  Thus, the spatial velocity of the remnant
star, $v_{\mathrm{spatial}} = \sqrt{v_{\mathrm{kick}}^2+v_{\mathrm{orb}}^2}$,
ranges from $\mathrm{108~km\,s^{-1}}$ (model ms\_20a) to $\mathrm{287~km\,s^{-1}}$ 
(model ms\_30a).  \citet{Ruiz04} showed that Tycho G star
has a spatial velocity of $\mathrm{136~km\,s^{-1}}$, which is located in the
range of spatial velocities in our models.

Furthermore, a surviving companion star will be strongly affected by
the impact of an SN~Ia and show distinguishing properties. The
identification of a surviving companion stars in historical supernova
remnants is a promising method to test progenitor models of 
SNe~Ia. Figure~\ref{Fig:remnant} shows the evolution of a remnant
companion star up to $5000\,\mathrm{s}$ after the supernova explosion for model
ms\_22a. The remnant companion star is puffed up due to the supernova
impact and heating, and it has a large radius of $\sim 4\,R_{\odot}$
(see the last snapshot of Fig.~\ref{Fig:remnant}). Although the
surviving companion star shrinks and relaxes to be almost spherical
again $5000\,\mathrm{s}$ after the explosion, it is out of thermal equilibrium and
its density and temperature are asymmetric (see
Fig.~\ref{Fig:remnant}). \citet{Pods03} modeled the post-impact
evolution of the surviving companion star and showed that the star is able
to completely recover thermal equilibrium $\sim 500\,\mathrm{yr}$ after the 
supernova explosion. He also pointed out that a surviving companion 
star may be significantly overluminous or underluminous $10^3-10^4\,\mathrm{yr}$ 
after the explosion relative to its pre-supernova luminosity
\citep{Pods03}. However, he did not simulate the dynamical interaction
of the companion star with the supernova ejecta. In our work, we
followed the detailed interaction of SNe~Ia and their companion
stars. It is necessary to carry out a separate calculation to follow
the post-impact evolution of a remnant star during its
re-equilibration phase, which will constrain the properties of the
surviving companion star for searches in historic SN~Ia remnants. This
will be addressed in a forthcoming study.

\section{Summary and Conclusions}
  \label{sec:conclusion}

  Based on a range of more realistic companion star models than those used
  in previous work, we have investigated the impact of Type Ia supernova ejecta on their
  companion stars in WD + MS binary systems using the SPH code
  GADGET3. As an initial model for our hydrodynamical impact simulations, all companion
  stars are constructed with Eggleton's stellar evolution code
  incorporating the possibility of an instability of the accretion
  disk around the WD and including the prescription of \citet{Hach99b}
  for mass accretion onto the WD. We summarize the 
  basic results and conclusions of our impact simulation as follows.

     \begin{enumerate}
     \item For our binary systems, we always find a stripped mass
       larger than $0.1\,M_{\odot}$. This is in disagreement with the
       most recent observational constraints on the detection of hydrogen
       in nebular spectra of SNe~Ia \citep{Leon07}.
     \item Such large stripped masses cause a serious problem for the
       SD scenario of SNe~Ia. However, prior mass-stripping from the
       companion star by an optically thick wind \citep{Hach99a, Hach08}
       or a spin-up spin-down phase as proposed by \citet{Just11, Di11, 
       Hach12} might be able to explain the absence of hydrogen 
       emission in nebular spectra of SNe~Ia.       
     \item For a given companion model, the dependence of stripped
       mass and kick velocity on the ratio of separation to the radius
       of the companion, $a_{\mathrm{f}}/R_{2}$, can be fitted by a power
       law. However, we do not find the same fit parameters as PRHW08
       when we put their models and ours together.  This indicates
       that details of the structure of companion stars are important for the results of
       the supernova impact.
     \item The differences to previous works are attributed to a
       more realistic treatment of the binary evolution of our
       progenitor models.
     \end{enumerate}

       Further improvements to the binary evolution and
       observational constraints for more SNe~Ia are
       needed, and more detailed modeling (in particular of the
       radiative transfer in the nebular phase) is required to
       determine under which circumstances the hydrogen will be
       detectable in observations.

\begin{acknowledgements}
      We thank the referee and the editor for their useful comments which
      helped to improve the paper. We thank Ken'ichi Nomoto, Achim Weiss  and Stefan Taubenberger for useful discussions. 
      Z.W.L and Z.W.H thank the financial support from the MPG-CAS Joint Doctoral 
      Promotion Program (DPP) and Max-Planck Institute for Astrophysics (MPA).
      This work is supported by the National Natural Science Foundation of China (Grant Nos.
      11033008 and 11103072),  the National Basic Research Program of China (Grant No. 2009CB824800) and 
      the Chinese Academy of Sciences (Grant N0. KJCX2-YW-T24).
      The work of F.K.R was supported by Deutsche Forschungsgemeinschaft via  the Emmy 
      Noether Program (RO 3676/1-1) and by the ARCHES prize of the German Federal 
      Ministry of Education and Research (BMBF). The simulations were carried out at 
      the Computing Center of the Max Plank Society, Garching,
      Germany.
\end{acknowledgements}

\bibliographystyle{aa}

\bibliography{zwliu-ref}

\begin{thebibliography}{76}
\expandafter\ifx\csname natexlab\endcsname\relax\def\natexlab#1{#1}\fi

\bibitem[{{Alexander} \& {Ferguson}(1994)}]{Alex94}
{Alexander}, D.~R. \& {Ferguson}, J.~W. 1994, \apj, 437, 879

\bibitem[{{Chen} \& {Tout}(2007)}]{Chen07}
{Chen}, X.-F. \& {Tout}, C.~A. 2007, \cjaa, 7, 245

\bibitem[{{Chomiuk} {et~al.}(2012){Chomiuk}, {Soderberg}, {Moe}, {Chevalier},
  {Rupen}, {Badenes}, {Margutti}, {Fransson}, {Fong}, \& {Dittmann}}]{Chom12}
{Chomiuk}, L., {Soderberg}, A.~M., {Moe}, M., {et~al.} 2012, \apj, 750, 164

\bibitem[{{Chugai}(1986)}]{Chug86}
{Chugai}, N.~N. 1986, \sovast, 30, 563

\bibitem[{{Di Stefano} {et~al.}(2011){Di Stefano}, {Voss}, \& {Claeys}}]{Di11}
{Di Stefano}, R., {Voss}, R., \& {Claeys}, J.~S.~W. 2011, \apjl, 738, L1+

\bibitem[{{Eggleton}(1971)}]{Eggl71}
{Eggleton}, P.~P. 1971, \mnras, 151, 351

\bibitem[{{Eggleton}(1972)}]{Eggl72}
{Eggleton}, P.~P. 1972, \mnras, 156, 361

\bibitem[{{Eggleton}(1973)}]{Eggl73}
{Eggleton}, P.~P. 1973, \mnras, 163, 279

\bibitem[{{Finzi} \& {Wolf}(1967)}]{Finz67}
{Finzi}, A. \& {Wolf}, R.~A. 1967, \apj, 150, 115

\bibitem[{{Geier} {et~al.}(2010){Geier}, {Heber}, {Kupfer}, \&
  {Napiwotzki}}]{Geie10}
{Geier}, S., {Heber}, U., {Kupfer}, T., \& {Napiwotzki}, R. 2010, \aap, 515,
  A37

\bibitem[{{Geier} {et~al.}(2007){Geier}, {Nesslinger}, {Heber}, {Przybilla},
  {Napiwotzki}, \& {Kudritzki}}]{Geie07}
{Geier}, S., {Nesslinger}, S., {Heber}, U., {et~al.} 2007, \aap, 464, 299

\bibitem[{{Hachisu} {et~al.}(1996){Hachisu}, {Kato}, \& {Nomoto}}]{Hach96}
{Hachisu}, I., {Kato}, M., \& {Nomoto}, K. 1996, \apjl, 470, L97+

\bibitem[{{Hachisu} {et~al.}(1999{\natexlab{a}}){Hachisu}, {Kato}, \&
  {Nomoto}}]{Hach99a}
{Hachisu}, I., {Kato}, M., \& {Nomoto}, K. 1999{\natexlab{a}}, \apj, 522, 487

\bibitem[{{Hachisu} {et~al.}(2008){Hachisu}, {Kato}, \& {Nomoto}}]{Hach08}
{Hachisu}, I., {Kato}, M., \& {Nomoto}, K. 2008, \apj, 679, 1390

\bibitem[{{Hachisu} {et~al.}(1999{\natexlab{b}}){Hachisu}, {Kato}, {Nomoto}, \&
  {Umeda}}]{Hach99b}
{Hachisu}, I., {Kato}, M., {Nomoto}, K., \& {Umeda}, H. 1999{\natexlab{b}},
  \apj, 519, 314

\bibitem[{{Hachisu} {et~al.}(2012){Hachisu}, {Kato}, {Saio}, \&
  {Nomoto}}]{Hach12}
{Hachisu}, I., {Kato}, M., {Saio}, H., \& {Nomoto}, K. 2012, \apj, 744, 69

\bibitem[{{Han} \& {Podsiadlowski}(2004)}]{Han04}
{Han}, Z. \& {Podsiadlowski}, P. 2004, \mnras, 350, 1301

\bibitem[{{Hillebrandt} \& {Niemeyer}(2000)}]{Hill00}
{Hillebrandt}, W. \& {Niemeyer}, J.~C. 2000, \araa, 38, 191

\bibitem[{{Horesh} {et~al.}(2012){Horesh}, {Kulkarni}, {Fox}, {Carpenter},
  {Kasliwal}, {Ofek}, {Quimby}, {Gal-Yam}, {Cenko}, {de Bruyn}, {Kamble},
  {Wijers}, {van der Horst}, {Kouveliotou}, {Podsiadlowski}, {Sullivan},
  {Maguire}, {Howell}, {Nugent}, {Gehrels}, {Law}, {Poznanski}, \&
  {Shara}}]{Hore12}
{Horesh}, A., {Kulkarni}, S.~R., {Fox}, D.~B., {et~al.} 2012, \apj, 746, 21

\bibitem[{{Hoyle} \& {Fowler}(1960)}]{Hoyl60}
{Hoyle}, F. \& {Fowler}, W.~A. 1960, \apj, 132, 565

\bibitem[{{Iben} \& {Tutukov}(1984)}]{Iben84}
{Iben}, Jr., I. \& {Tutukov}, A.~V. 1984, \apj, 284, 719

\bibitem[{{Iglesias} \& {Rogers}(1996)}]{Igle96}
{Iglesias}, C.~A. \& {Rogers}, F.~J. 1996, \apj, 464, 943

\bibitem[{{Ivanova} \& {Taam}(2004)}]{Ivan04}
{Ivanova}, N. \& {Taam}, R.~E. 2004, \apj, 601, 1058

\bibitem[{{Justham}(2011)}]{Just11}
{Justham}, S. 2011, \apjl, 730, L34+

\bibitem[{{Kasen}(2010)}]{Kase10}
{Kasen}, D. 2010, \apj, 708, 1025

\bibitem[{{Kato}(2011)}]{Kato11}
{Kato}, M. 2011, arXiv:astro-ph/1110.0055

\bibitem[{{Kato} \& {Hachisu}(2004)}]{Kato04}
{Kato}, M. \& {Hachisu}, I. 2004, \apjl, 613, L129

\bibitem[{{Kerzendorf} {et~al.}(2009){Kerzendorf}, {Schmidt}, {Asplund},
  {Nomoto}, {Podsiadlowski}, {Frebel}, {Fesen}, \& {Yong}}]{Kerz09}
{Kerzendorf}, W.~E., {Schmidt}, B.~P., {Asplund}, M., {et~al.} 2009, \apj, 701,
  1665

\bibitem[{{Lasota}(2001)}]{Laso01}
{Lasota}, J.-P. 2001, \nar, 45, 449

\bibitem[{{Leibundgut}(2008)}]{Leib08}
{Leibundgut}, B. 2008, General Relativity and Gravitation, 40, 221

\bibitem[{{Leonard}(2007)}]{Leon07}
{Leonard}, D.~C. 2007, \apj, 670, 1275

\bibitem[{{Mannucci}(2005)}]{Mann05}
{Mannucci}, F. 2005, in Astronomical Society of the Pacific Conference Series,
  Vol. 342, 1604-2004: Supernovae as Cosmological Lighthouses, ed. {M.~Turatto,
  S.~Benetti, L.~Zampieri, \& W.~Shea}, 140

\bibitem[{{Mannucci}(2009)}]{Mann09}
{Mannucci}, F. 2009, in American Institute of Physics Conference Series, Vol.
  1111, American Institute of Physics Conference Series, ed. {G.~Giobbi,
  A.~Tornambe, G.~Raimondo, M.~Limongi, L.~A.~Antonelli, N.~Menci, \&
  E.~Brocato}, 467--476

\bibitem[{{Maoz} \& {Badenes}(2010)}]{Maoz10a}
{Maoz}, D. \& {Badenes}, C. 2010, \mnras, 407, 1314

\bibitem[{{Maoz} \& {Mannucci}(2011)}]{Maoz11}
{Maoz}, D. \& {Mannucci}, F. 2011, arXiv:astro-ph/1111.4492

\bibitem[{{Maoz} {et~al.}(2010){Maoz}, {Sharon}, \& {Gal-Yam}}]{Maoz10b}
{Maoz}, D., {Sharon}, K., \& {Gal-Yam}, A. 2010, \apj, 722, 1879

\bibitem[{{Marietta} {et~al.}(2000){Marietta}, {Burrows}, \&
  {Fryxell}}]{Mari00}
{Marietta}, E., {Burrows}, A., \& {Fryxell}, B. 2000, \apjs, 128, 615

\bibitem[{{Mattila} {et~al.}(2005){Mattila}, {Lundqvist}, {Sollerman}, {Kozma},
  {Baron}, {Fransson}, {Leibundgut}, \& {Nomoto}}]{Matt05}
{Mattila}, S., {Lundqvist}, P., {Sollerman}, J., {et~al.} 2005, \aap, 443, 649

\bibitem[{{Meng} {et~al.}(2007){Meng}, {Chen}, \& {Han}}]{Meng07}
{Meng}, X., {Chen}, X., \& {Han}, Z. 2007, \pasj, 59, 835

\bibitem[{{Meng} \& {Yang}(2010)}]{Meng10}
{Meng}, X. \& {Yang}, W. 2010, \apj, 710, 1310

\bibitem[{{Mennekens} {et~al.}(2010){Mennekens}, {Vanbeveren}, {De Greve}, \&
  {De Donder}}]{Menn10}
{Mennekens}, N., {Vanbeveren}, D., {De Greve}, J.~P., \& {De Donder}, E. 2010,
  \aap, 515, A89+

\bibitem[{{Nelemans} {et~al.}(2005){Nelemans}, {Napiwotzki}, {Karl}, {Marsh},
  {Voss}, {Roelofs}, {Izzard}, {Montgomery}, {Reerink}, {Christlieb}, \&
  {Reimers}}]{Nele05}
{Nelemans}, G., {Napiwotzki}, R., {Karl}, C., {et~al.} 2005, \aap, 440, 1087

\bibitem[{{Nomoto}(1982)}]{Nomo82}
{Nomoto}, K. 1982, \apj, 253, 798

\bibitem[{{Nomoto} \& {Iben}(1985)}]{Nomo85}
{Nomoto}, K. \& {Iben}, Jr., I. 1985, \apj, 297, 531

\bibitem[{{Nomoto} {et~al.}(1984){Nomoto}, {Thielemann}, \& {Yokoi}}]{Nomo84}
{Nomoto}, K., {Thielemann}, F.-K., \& {Yokoi}, K. 1984, \apj, 286, 644

\bibitem[{{Pakmor} {et~al.}(2012{\natexlab{a}}){Pakmor}, {Edelmann},
  {R{\"o}pke}, \& {Hillebrandt}}]{Pakm12a}
{Pakmor}, R., {Edelmann}, P., {R{\"o}pke}, F.~K., \& {Hillebrandt}, W.
  2012{\natexlab{a}}, \mnras, 424, 2222

\bibitem[{{Pakmor} {et~al.}(2011){Pakmor}, {Hachinger}, {R{\"o}pke}, \&
  {Hillebrandt}}]{Pakm11b}
{Pakmor}, R., {Hachinger}, S., {R{\"o}pke}, F.~K., \& {Hillebrandt}, W. 2011,
  \aap, 528, A117+

\bibitem[{{Pakmor} {et~al.}(2010){Pakmor}, {Kromer}, {R{\"o}pke}, {Sim},
  {Ruiter}, \& {Hillebrandt}}]{Pakm10}
{Pakmor}, R., {Kromer}, M., {R{\"o}pke}, F.~K., {et~al.} 2010, \nat, 463, 61

\bibitem[{{Pakmor} {et~al.}(2012{\natexlab{b}}){Pakmor}, {Kromer},
  {Taubenberger}, {Sim}, {R{\"o}pke}, \& {Hillebrandt}}]{Pakm12b}
{Pakmor}, R., {Kromer}, M., {Taubenberger}, S., {et~al.} 2012{\natexlab{b}},
  \apjl, 747, L10

\bibitem[{{Pakmor} {et~al.}(2008){Pakmor}, {R{\"o}pke}, {Weiss}, \&
  {Hillebrandt}}]{Pakm08}
{Pakmor}, R., {R{\"o}pke}, F.~K., {Weiss}, A., \& {Hillebrandt}, W. 2008, \aap,
  489, 943

\bibitem[{{Pan} {et~al.}(2010){Pan}, {Ricker}, \& {Taam}}]{Pan10}
{Pan}, K.-C., {Ricker}, P.~M., \& {Taam}, R.~E. 2010, \apj, 715, 78

\bibitem[{{Pan} {et~al.}(2012){Pan}, {Ricker}, \& {Taam}}]{Pan12}
{Pan}, K.-C., {Ricker}, P.~M., \& {Taam}, R.~E. 2012, \apj, 750, 151

\bibitem[{{Patat} {et~al.}(2007){Patat}, {Chandra}, {Chevalier}, {Justham},
  {Podsiadlowski}, {Wolf}, {Gal-Yam}, {Pasquini}, {Crawford}, {Mazzali},
  {Pauldrach}, {Nomoto}, {Benetti}, {Cappellaro}, {Elias-Rosa}, {Hillebrandt},
  {Leonard}, {Pastorello}, {Renzini}, {Sabbadin}, {Simon}, \&
  {Turatto}}]{Pata07}
{Patat}, F., {Chandra}, P., {Chevalier}, R., {et~al.} 2007, Science, 317, 924

\bibitem[{{Paxton} {et~al.}(2011){Paxton}, {Bildsten}, {Dotter}, {Herwig},
  {Lesaffre}, \& {Timmes}}]{Paxt11}
{Paxton}, B., {Bildsten}, L., {Dotter}, A., {et~al.} 2011, \apjs, 192, 3

\bibitem[{{Perlmutter} {et~al.}(1999){Perlmutter}, {Aldering}, {Goldhaber},
  {Knop}, {Nugent}, {Castro}, {Deustua}, {Fabbro}, {Goobar}, {Groom}, {Hook},
  {Kim}, {Kim}, {Lee}, {Nunes}, {Pain}, {Pennypacker}, {Quimby}, {Lidman},
  {Ellis}, {Irwin}, {McMahon}, {Ruiz-Lapuente}, {Walton}, {Schaefer}, {Boyle},
  {Filippenko}, {Matheson}, {Fruchter}, {Panagia}, {Newberg}, {Couch}, \& {The
  Supernova Cosmology Project}}]{Perl99}
{Perlmutter}, S., {Aldering}, G., {Goldhaber}, G., {et~al.} 1999, \apj, 517,
  565

\bibitem[{{Phillips}(1993)}]{Phil93}
{Phillips}, M.~M. 1993, \apjl, 413, L105

\bibitem[{{Phillips} {et~al.}(1999){Phillips}, {Lira}, {Suntzeff}, {Schommer},
  {Hamuy}, \& {Maza}}]{Phil99}
{Phillips}, M.~M., {Lira}, P., {Suntzeff}, N.~B., {et~al.} 1999, \aj, 118, 1766

\bibitem[{{Podsiadlowski}(2003)}]{Pods03}
{Podsiadlowski}, P. 2003, arXiv:astro-ph/0303660

\bibitem[{{Pols} {et~al.}(1997){Pols}, {Tout}, {Schroder}, {Eggleton}, \&
  {Manners}}]{Pols97}
{Pols}, O.~R., {Tout}, C.~A., {Schroder}, K.-P., {Eggleton}, P.~P., \&
  {Manners}, J. 1997, \mnras, 289, 869

\bibitem[{{Riess} {et~al.}(1998){Riess}, {Filippenko}, {Challis},
  {Clocchiatti}, {Diercks}, {Garnavich}, {Gilliland}, {Hogan}, {Jha},
  {Kirshner}, {Leibundgut}, {Phillips}, {Reiss}, {Schmidt}, {Schommer},
  {Smith}, {Spyromilio}, {Stubbs}, {Suntzeff}, \& {Tonry}}]{Ries98}
{Riess}, A.~G., {Filippenko}, A.~V., {Challis}, P., {et~al.} 1998, \aj, 116,
  1009

\bibitem[{{Rodr{\'{\i}}guez-Gil} {et~al.}(2010){Rodr{\'{\i}}guez-Gil},
  {Santander-Garc{\'{\i}}a}, {Knigge}, {Corradi}, {G{\"a}nsicke}, {Barlow},
  {Drake}, {Drew}, {Miszalski}, {Napiwotzki}, {Steeghs}, {Wesson}, {Zijlstra},
  {Jones}, {Liimets}, {Mu{\~n}oz-Darias}, {Pyrzas}, \&
  {Rubio-D{\'{\i}}ez}}]{Rodr10}
{Rodr{\'{\i}}guez-Gil}, P., {Santander-Garc{\'{\i}}a}, M., {Knigge}, C.,
  {et~al.} 2010, \mnras, 407, L21

\bibitem[{{Ruiter} {et~al.}(2009){Ruiter}, {Belczynski}, \& {Fryer}}]{Ruit09}
{Ruiter}, A.~J., {Belczynski}, K., \& {Fryer}, C. 2009, \apj, 699, 2026

\bibitem[{{Ruiz-Lapuente} {et~al.}(2004){Ruiz-Lapuente}, {Comeron},
  {M{\'e}ndez}, {Canal}, {Smartt}, {Filippenko}, {Kurucz}, {Chornock}, {Foley},
  {Stanishev}, \& {Ibata}}]{Ruiz04}
{Ruiz-Lapuente}, P., {Comeron}, F., {M{\'e}ndez}, J., {et~al.} 2004, \nat, 431,
  1069

\bibitem[{{Saio} \& {Nomoto}(1998)}]{Saio98}
{Saio}, H. \& {Nomoto}, K. 1998, \apj, 500, 388

\bibitem[{{Schaefer} \& {Pagnotta}(2012)}]{Scha12}
{Schaefer}, B.~E. \& {Pagnotta}, A. 2012, \nat, 481, 164

\bibitem[{{Schroder} {et~al.}(1997){Schroder}, {Pols}, \& {Eggleton}}]{Schr97}
{Schroder}, K.-P., {Pols}, O.~R., \& {Eggleton}, P.~P. 1997, \mnras, 285, 696

\bibitem[{{Springel}(2005)}]{Spri05}
{Springel}, V. 2005, \mnras, 364, 1105

\bibitem[{{Springel} {et~al.}(2001){Springel}, {Yoshida}, \& {White}}]{Spri01}
{Springel}, V., {Yoshida}, N., \& {White}, S.~D.~M. 2001, \na, 6, 79

\bibitem[{{Sternberg} {et~al.}(2011){Sternberg}, {Gal-Yam}, {Simon}, {Leonard},
  {Quimby}, {Phillips}, {Morrell}, {Thompson}, {Ivans}, {Marshall},
  {Filippenko}, {Marcy}, {Bloom}, {Patat}, {Foley}, {Yong}, {Penprase},
  {Beeler}, {Allende Prieto}, \& {Stringfellow}}]{Ster11}
{Sternberg}, A., {Gal-Yam}, A., {Simon}, J.~D., {et~al.} 2011, Science, 333,
  856

\bibitem[{{van Paradijs}(1996)}]{van96}
{van Paradijs}, J. 1996, \apjl, 464, L139+

\bibitem[{{Wang} {et~al.}(2009){Wang}, {Chen}, {Meng}, \& {Han}}]{Wang09}
{Wang}, B., {Chen}, X., {Meng}, X., \& {Han}, Z. 2009, \apj, 701, 1540

\bibitem[{{Wang} \& {Han}(2010)}]{Wang10a}
{Wang}, B. \& {Han}, Z. 2010, \aap, 515, A88+

\bibitem[{{Wang} {et~al.}(2010){Wang}, {Li}, \& {Han}}]{Wang10b}
{Wang}, B., {Li}, X.-D., \& {Han}, Z.-W. 2010, \mnras, 401, 2729

\bibitem[{{Webbink}(1984)}]{Webb84}
{Webbink}, R.~F. 1984, \apj, 277, 355

\bibitem[{{Wheeler} {et~al.}(1975){Wheeler}, {Lecar}, \& {McKee}}]{Whee75}
{Wheeler}, J.~C., {Lecar}, M., \& {McKee}, C.~F. 1975, \apj, 200, 145

\bibitem[{{Whelan} \& {Iben}(1973)}]{Whel73}
{Whelan}, J. \& {Iben}, Jr., I. 1973, \apj, 186, 1007

\end{thebibliography}

\end{document}